\documentclass[iop]{emulateapj}
\usepackage{natbib}
\usepackage{txfonts}
\usepackage{graphicx}
\usepackage{wasysym}

\newcommand{\mub}{$\mu_{B}$}
\newcommand{\muv}{$\mu_{V}$}
\newcommand{\bmv}{B$-$V}
\newcommand{\feh}{[Fe/H]}

\newcommand{\sfrone}{$M_{\astrosun}$~yr$^{-1}$}
\newcommand{\sfrtwo}{$M_{\astrosun}$~yr$^{-1}$~kpc$^{-2}$}

\shortauthors{Watkins et al.}

\begin{document}

\title{The Red and Featureless Outer Disks of Nearby Spiral Galaxies}
\shorttitle{The Outer Disks of Nearby Spiral Galaxies}

\author{Aaron E. Watkins\altaffilmark{1}, 
  J. Christopher Mihos\altaffilmark{1},
  Paul Harding\altaffilmark{1}
  }

\altaffiltext{1}{Department of Astronomy, Case Western Reserve
  University, Cleveland, OH 44106, USA}

\begin{abstract}

We present results from deep, wide-field surface photometry of three
nearby (D=4--7 Mpc) spiral galaxies: M94 (NGC~4736), M64 (NGC~4826),
and M106 (NGC~4258). Our imaging reaches limiting surface
brightnesses of \mub $\sim$ 28 -- 30 mag arcsec$^{-2}$ and probes
colors down to \mub $\sim$ 27.5 mag arcsec$^{-2}$. We compare our
broadband optical data to available ultraviolet and high
column-density \ion{H}{1} data to better constrain the star forming
history and stellar populations of the outermost parts of each
galaxy's disk. Each galaxy has a well-defined radius beyond which
little star formation occurs and the disk light appears both
azimuthally smooth and red in color, suggestive of old, well-mixed
stellar populations. Given the lack of ongoing star formation or blue
stellar populations in these galaxies' outer disks, the most likely
mechanisms for their formation are dynamical processes such as disk
heating or radial migration, rather than inside-out growth of the
disks. This is also implied by the similarity in outer disk properties
despite each galaxy showing distinct levels of environmental
influence, from a purely isolated galaxy (M94) to one experiencing
weak tidal perturbations from its satellite galaxies (M106) to a
galaxy recovering from a recent merger (M64), suggesting that a
variety of evolutionary histories can yield similar outer disk
structure.  While this suggests a common secular mechanism for outer
disk formation, the large extent of these smooth, red stellar
populations---which reach several disk scalelengths beyond the
galaxies' spiral structure---may challenge models of radial migration
given the lack of any non-axisymmetric forcing at such large radii.

\end{abstract}

\keywords{galaxies:individual(M94,M64,M106), galaxies:evolution,
  galaxies:spiral}

\newpage

\section{Introduction}

Disk galaxies present some of the cleanest laboratories to test
theories of galaxy formation and evolution. At first glance, their
stellar populations appear distinctly segregated, both spatially and
kinematically, with the older, kinematically hotter stars forming
central bulges or bars, and the younger, kinematically colder stars
forming disks around these bulges. In the hierarchical accretion model
of galaxy formation \citep[e.g.][]{searle78, wf91, springel05,
  vogelsberger14}, this structure is explained by an ``inside-out''
formation mechanism, leading to the prediction that the mean age of
the stellar populations will be lowest in the disk outskirts. Stars
near the galaxy center, formed from primordial gas early in the
universe's history, would quickly enrich their local ISM
\citep{mcclure69, wyse92}, while in the outskirts the lower gas
densities and star formation rates result in reduced enrichment
efficiencies \citep{schmidt59, kennicutt98, bigiel08,
  krumholz12}. Thus, there should exist separate, well-defined
age-metallicity relationships (AMRs) at individual radii throughout
the disk, with an overall negative radial metallicity gradient for
stellar populations of a given age \citep{twarog80, chippini97,
  naab06}. Beyond a certain radius \citep[when the gas density falls
  below a ``critical'' value;][]{kennicutt89}, one would expect to
find very few, if any, stars formed {\it in-situ}.

This simple and elegant picture, however, is not fully supported by
observation. Measurements of the AMR in the solar neighborhood, for
example, have shown a much larger dispersion than is expected under
such a model \citep{edvardsson93, haywood06}. The Milky Way's
metallicity gradient also appears to flatten beyond $\sim10$kpc
\citep[e.g.][but see Luck \& Lambert 2011, Lemasle et
  al. 2013]{twarog97, yong05, maciel10}; other disk galaxies seem to
show similar behavior \citep[e.g.][]{bresolin09, vlajic09, vlajic11,
  sanchez14}. Also, while many disks show negative age gradients in
their stellar populations \citep[e.g.][]{sanchez14}, this trend may
often reverse beyond a certain radius \citep[e.g.][]{bakos08,
  zheng15}. Finally, disk stars are found to inhabit the very extended
outer reaches of the host galaxies, sometimes beyond the apparent star
formation threshold radius \citep[e.g.][]{tiede04, davidge06,
  davidge07, azzollini08, bakos08, vlajic09, martin14, okamoto15,
  zheng15}, including patches of new star formation
\citep[e.g.][]{thilker05, gildepaz05, thilker07, lemonias11,
  yildiz15}. Under the simple model described above, the presence of
these stars is mystifying.

Reconciling these inconsistencies with theory has prompted detailed
investigation into the inner workings of disk galaxies, and some amount
of consensus is beginning to emerge. Radial migration of stars appears
to be important, wherein stars move radially throughout the disk via
resonances with inner bars or spiral arms, while maintaining their
reasonably circular orbits \citep{sellwood02, debattista06}. Numerous
simulations have shown that this process can move stars in excess of a
kiloparsec from their birthplaces \citep{sellwood02, roskar08,
schonrich09, sanchez09, minchev11, minchev13}, effectively flattening
the AMR \citep{schonrich09, sanchez09, minchev13}. Additionally,
migration can serve as a means of growing the disk radially even in the
presence of a star formation threshold \citep{roskar08, sanchez09,
roskar12, minchev12, minchev13}. While the detailed mechnisms differ in
the simulations, a common prediction is the creation of a ``U-shaped''
age gradient in the stellar populations, where older disk stars are found
inhabiting both the inner regions of the galaxy and its extended
outer disk. Coupled with a flat metallicity gradient in the disk
periphery, this would give rise to the ``U-shaped'' radial color profile
observed in most optical bands. In the simulations by \citet{roskar08},
the break in the age gradient is always coupled to a break in the radial
mass and luminosity surface density profiles, and is pushed farther out
over time as the stellar disk gains mass \citep[but see][]{sanchez09}.

Increasingly, observations appear to support this picture.
\citet{bakos08}, using galaxies in the frequently-observed Stripe 82
of the Sloan Digital Sky Survey \citep[SDSS;][]{york00}, found that
this predicted ``U-shaped'' color profile is in fact quite common in
disk galaxies, and is also commonly coupled to a so-called ``Type II''
\citep[or downbending: see][]{pohlen06, erwin08} break in surface
brightness, in apparent agreement with the simulations of
\citet{roskar08}.  However, they found no such break in the mass
surface density, in better agreement with \citet{sanchez09}.
Follow-up studies showed similar behavior \citep{bakos08, gutierrez11,
  laine14, martin14}.  Studies of outer disks using resolved stars
also frequently reveal a shorter scale-length for younger, main
sequence populations than for red giant branch (RGB) stars
\citep[e.g.][]{davidge06, vlajic09, radburn11a}, again implying outer
disk populations dominated by old stars, although a break in surface
brightness is not always present \citep[as in the case of NGC 300 and
  NGC 7793;][]{bland05, vlajic09, vlajic11}.

This schema of radial migration hence appears to be mostly sound, yet
exceptions do exist to complicate it. One obvious example is the
existence of \citep[again using the nomenclature of][]{erwin08} ``Type
I'' disks (a constant slope exponential surface brightness profile)
and ``Type III'' disks (an upbending break), both of which often show
flat color profiles, at a relatively bluer color than the downbending
``Type II'' disks \citep{zheng15}. If the model proposed by
\citet{roskar08} is universal, such galaxies would require other
mechanisms to shape their disks, such as the accretion of low mass
satellites \citep[e.g.][]{younger07}. However, while Type III disks do
show evidence of environmental influences \citep{laine14}, the
correlation between environment and Type I and Type II disk fractions
is unclear \citep[e.g.][]{pohlen06}.

Extended ultraviolet (XUV) disks \citep{thilker07} also pose intriguing
questions regarding the assumption of a star formation threshold, a
seemingly necessary aspect of the models described above.
\citet{zaritsky07} used GALEX imaging to show that that perhaps as many
as 50\% of disk galaxies contain young ($<400$ Myr) star clusters
between 1.25$R_{25}$ and 2$R_{25}$ (out to surface brightnesses of
roughly \muv$\sim$29), which, given their number density and assuming a
constant star formation history over 10 Gyrs, could fully account for
the measured surface brightnesses of starlight at these radii. This
would imply that radial migration may not be universally necessary to
build outer disks. Indeed, a recent spectroscopic study by
\citet{morelli15} found mostly negative metallicity gradients in the
oldest stellar populations in their sample of disk galaxies, difficult
to explain in the context of significant radial migration. Additionally,
disk metallicity gradients do not appear to be affected by the presence
or absence of bars \citep{sanchez14}, in apparent contradiction to the
simulations of \citet{minchev11, minchev12, minchev13}.

Satellite interactions can also drive evolution in the properties of
disks. These interactions can transfer angular momentum to disk stars
either directly from the satellite companion \citep[e.g.][]{walker96}
or from disk gas driven inwards by the effects of the interaction
\citep[e.g.][]{hernquist95}, leading to a radial spreading of the disk
\citep{younger07}. Tidal stripping of the satellite companion can also
deposit stars in the outer disk of the host as well
\citep[e.g.][]{stewart09}. Tidal encounters may also induce localized
star formation in the disk outskirts \citep[e.g.][]{whitmore95,
  weilbacher00, smith08, powell13}, or potentially generate extremely
extended spiral arms \citep{koribalski09, khoperskov15}, which may
then form new stars and drive migration of older stars further
outwards in the disk. The affects of accretion and tidal interaction,
however, depend strongly on the orbital parameters of the encounter
\citep[e.g.  mass ratios of the interacting galaxies, prograde
  vs. retrograde orbits, etc.;][]{toomre72, barnes88, quinn93,
  walker96, bournaud05, donghia10}. This, in combination with
progenitors with potentially different structural properties, star
formation histories, and metallicity distributions, implies that the
influence of accretion and interaction events on disk galaxies ought
to be stochastic in nature.

A common thread amongst most of the observational studies cited above,
save for those using star counts, is the use of azimuthal averaging
when constructing one-dimensional radial profiles of the galaxy
light. Such studies measure the surface brightness or color of the
disk in successively larger radial bins, thereby maintaining a high
signal-to-noise ratio even in the faint outer isophotes of the galaxy.
This method has proven extremely useful for large statistical studies
\citep[e.g.][]{pohlen06, erwin08, bakos08, martin14, zheng15}, but it
suffers from a number of pitfalls when applied on a galaxy-by-galaxy
basis. These techniques ``average out'' azimuthal asymmetries in the
surface brightness and color of the outer disk, which often hold
important clues about its dynamical history \citep[see, e.g., the case
  of M101][]{mihos13a}. These asymmetries may also skew the results of
azimuthal averaging by mixing disk light with regions of blank
background sky, a particular problem when radial bins with constant
ellipticity and position angle are used at all radii \citep[which is
  often the case;][]{pohlen06, erwin08}. In some cases, inferences
drawn from azimuthal averaging may even depend on the choice of metric
used to construct the profile. For example, in the outskirts of an XUV
disk, a luminosity-weighted average surface brightness will present a
much different story than an areal-weighted median surface brightness,
as most of the light in the outer disk will be contained in just a few
blue pockets of star formation \citep[e.g.  M83;][]{thilker05}. The
exact importance of these various pitfalls to azimuthal averaging is
not yet clear.

Given these complications, more detailed studies of individual
galaxies may provide important new tests for the current paradigm of
disk galaxy evolution.  For example, if weak spiral arms persist
beyond the so-called truncation radius \citep[e.g.][]{khoperskov15},
this may drive outer disk star formation \citep{bush08} and lead to
the radial growth of disks with time; such features may be washed out
by an azimuthally-averaged photometric analysis.  Bright, nearby disk
galaxies provide the best targets for such work; their proximity
allows us to study them at high spatial resolution, and also permits
follow-up studies of their discrete stellar populations.  While many
spatially resolved studies have been done at high surface brightness
(\mub $\lesssim$ 26) in the past \citep[for just a few examples,
  see][]{schweizer76, okamura78, yuan81, kennicutt86, tacconi90},
recent improvements in deep imaging techniques now allow us to probe
the outer disks of these galaxies using similar techniques down to the
much lower surface brightnesses characteristic of their extreme outer
disks.

Here we present deep surface photometry of three large nearby disk
galaxies --- M106, M94, and M64 --- to explore the structure and
stellar populations in their outer disks. Taken using the Burrell
Schmidt Telescope at Kitt Peak National Observatory (KPNO), our data
reach limiting surface brightnesses of \mub$\sim$ 28--30 mags
arcsec$^{-2}$ in B and V, and we combine our data with extant GALEX
and 21cm neutral hydrogen maps of each system to explore the efficacy
of different formation mechanisms for outer disks.  In Section 2, we
present our observation and data reduction strategies; in Section 3,
we discuss our methods for extracting and analyzing the surface
brightness and color profiles of the galaxies; in Section 4 we present
and discuss our results on a galaxy-by-galaxy basis; in Section 5 we
discuss the implications of these results in the context of galactic
evolution; and in Section 6 we present a summary of our results and
conclusions.

\section{Observational Data}

\subsection{Deep Optical Imaging}

We obtained deep broadband imaging of the galaxies M64, M94, and M106
using CWRU's Burrell Schmidt Telescope at KPNO on moonless,
photometric nights in Spring 2012 and Spring 2013. Our observing
strategy and data reduction techniques are described in detail in
\citep[and references therein]{watkins14}, and we repeat only the most
important details here.  The telescope's field of view is
$1.65^{\circ} \times 1.65^{\circ}$, imaged onto a $4096 \times 4096$
STA0500A CCD, for a pixel scale of $1.45$\arcsec\ pixel$^{-1}$. We
observed in two filters: a modified Johnson \emph{B} (2012), and
Washington \emph{M} (2013). The latter filter is a proxy for Johnson
\emph{V}; it is similar in width but $\sim$ 200\AA\ bluer, and
effectively cuts out diffuse airglow from the bright \ion{O}{1}
$\lambda$5577 line \citep{feldmeier02}. Each exposure was 1200s in
\emph{B} and 900s in \emph{M}, with $\sim$0.5$^{\circ}$ dithers
between exposures to reduce contamination from large-scale artifacts
such as scattered light and flat-fielding errors. For each galaxy, the
total exposure times are as follows: for M94, $24\times1200$s
(\emph{B}) and $32\times900$s (\emph{M}); for M64, $23\times1200$s
(\emph{B}) and $30\times900$s (\emph{M}); for M106, $27\times1200$s
(\emph{B}) and $38\times900$s (\emph{M}). Sky levels in each exposure
were 700 -- 900 ADU pixel$^{-1}$ in \emph{B} and 1200 -- 1400 ADU
pixel$^{-1}$ in \emph{M}.

In addition to the object frames, we also observed offset blank sky
pointings for use in constructing night sky flats. We alternated between
observing object frames and blank sky frames, in order to maintain
similar observing conditions between the two and minimize systematic
differences due to changes in telescope flexure and night sky
conditions. However, during data reduction we found that the only
measurable difference in flat fields constructed from the various
subsets of sky frames (taken object by object or run by run) was a mild
seasonal gradient that was easily corrected for \citep[for details, see
Section 2.2 in][]{watkins14}. Thus, in the end we constructed our final
sky flat using all sky exposures taken throughout each observing season,
resulting in $\sim$100 sky frames in \emph{B} and $\sim$120 in
\emph{M}.

Finally, during each season, we also observed Landolt standard fields
\citep{landolt92} to derive color terms for each filter, along with deep
images of Procyon (1200s in \emph{B}) and Regulus (900s in \emph{M}) to
measure the extended point spread function and characterize reflections
between the CCD, dewar window, and filter \citep{slater09}.

We begin the data reduction by first applying standard overscan and
bias subtraction, then correcting for nonlinear chip response and
adding a world coordinate system (WCS) to each image. We derive
photometric zeropoints for each image using Sloan Digital Sky Survey
\citep[SDSS, DR8;][]{aihara11} stars located in the field, converting
their \emph{ugriz} magnitudes to Johnson \emph{B} and \emph{V} by
adopting the prescription of \citet{lupton05} and only using stars
within the color range \bmv$=0$ to $1.5$. We use these zeropoints and
the color terms derived from the Landolt standard stars to convert our
magnitudes into standard Johnson \emph{B} and \emph{V} magnitudes,
which we use in all of our analyses throughout this paper. In our
final mosaics, we are able to recover converted SDSS magnitudes of
SDSS stars in-frame to $\sigma_{V}=0.03$ and $\sigma_{B-V}=0.04$ for
all three galaxies. These are hence the {\it absolute} photometric
uncertainties on any magnitudes and colors we quote in this paper; it
should be noted that these include the intrinsic scatter in both the
transformation between SDSS and Johnson photometric systems, and in
the transformation from our custom filters to the Johnson
system. However, {\it relative} photometric uncertainties within a
single mosaic are typically much lower than this at high surface
brightness ($\sigma_{V}<0.01$); at low surface brightness, the
relative photometric uncertainty is dominated by uncertainty in the
sky-subtracted background.  In each mosaic, this background uncertainty
(in the vicinity of each galaxy) is typically of order $\pm 1$ADU
($\sim 0.1$\% of sky; see above), which implies a global limiting
surface brightness of $\mu_{B,lim} \sim 29.5$, although local limiting
surface brightnesses vary across each mosaic. The limiting surface
brightness in the mosaic of M64 is significantly brighter than the other
two ($\mu_{B,lim} \sim 28.0$) due to the presence of foreground
Galactic cirrus; we discuss this in more depth in Section 4.2.

We constructed flat fields in each filter using the offset night sky
frames. For each sky image, we applied an initial mask using
IRAF's\footnote{IRAF is distributed by the National Optical Astronomy
Observatory, which is operated by the Association of Universities for
Research in Astronomy (AURA), Inc., under cooperative agreement with the
National Science Foundation.} \emph{objmask} task, hand-masked any
diffuse light missed by \emph{objmask} (typically scattered light from
stars located just off frame), and combined the images into a
preliminary flat. We then flattened each sky frame using this
preliminary flat, modeled and subtracted sky planes from each flattened
image, and created a new flat from the sky-subtracted images. We
repeated this step five times, at which point the resulting flat-field
converged. We then corrected these master flats for the seasonal
residual planes described above before applying them to the images.

The last steps of the reduction process consist of star and sky
subtraction, followed by final mosaicing. We first subtract the
diffuse halos around bright stars following the technique of
\citet{slater09}.  These halos arise from both the extended stellar
PSF and from reflections between the CCD, dewar window, and filter. We
create a model for these halos by measuring them from our deep imaging
of Procyon and Regulus, then scale and subtract these models from each
star brighter than \emph{V}$=$10.5 in each frame. We then mask each
image of all bright stars and galaxies and fit planes to the diffuse
night sky background. After subtracting these planes from each image,
the images are ready to be combined into the final mosaics. We use
IRAF's \emph{wregister} and \emph{imcombine} tasks to create these
mosaics, using a median combine after scaling each image to a common
photometric zeropoint. Full width half max (FWHM) values of the
stellar point spread functions (PSFs) are nearly the same on all
mosaics: $\sim$2.2 pixels, or 3.2\arcsec.  This large value is a
combination of registration error and seeing variations throughout the
observing runs; FWHM on individual exposures is typically much smaller
($\sim$1.5\arcsec \ - \ 2\arcsec).

Once the final mosaics are complete, we also create masked and rebinned
versions to improve signal-to-noise at low surface brightness and better
reveal faint extended features in the outer regions of each galaxy. The
masking process typically only masks pixels brighter than $\mu_{B}
\approx 27$ and $\mu_{V} \approx 26$, and leaves fainter pixels
untouched. After masking, we rebin the mosaic into $9 \times 9$ pixel
(13\arcsec $\times$ 13\arcsec) blocks and calculate the median in each
block to create these ``low surface brightness enhanced'' mosaics.

\subsection{Ancillary Multi-Wavelength Datasets}

To supplement our broadband imaging, and to study the star forming
properties and neutral hydrogen distribution in each of our galaxies, we
have obtained ancillary ultraviolet and 21-cm radio data from a variety
of sources.

The ultraviolet data comes from several different surveys done by the
Galaxy Evolution Explorer \citep[GALEX;][]{martin05} mission, downloaded
from the GR6/GR7 data release\footnote{http://galex.stsci.edu/GR6/}. The
far ultraviolet (FUV; 1350-1750 \AA) emission traces recent ($<50$ Myr)
star formation, while near ultraviolet (NUV; 1750-2750 \AA) emission
traces slightly older ($<100$ Myr) populations, along with some
contribution from evolved horizontal branch populations. We use the
deepest available images for each galaxy, which come from different
surveys; for M94, we used the Nearby Galaxies Survey
\citep[NGS;][]{bianchi03}; for M64, we used the Calibration Imaging
Survey (CIS), in which it appears serendipitously at the edge of the
field containing the white dwarf WDST\_GD\_153\_0003 (as such, the M64
UV data is the shallowest); for M106, we used the Deep Imaging Survey
(DIS).

Additionally, we obtained \ion{H}{1} data from The \ion{H}{1} Nearby
Galaxies Survey \citep[THINGS;][]{walter08} for M94 and M64, and from
the Westerbork \ion{H}{1} Survey of Irregular and SPiral Galaxies
\citep[WHISP;][]{vanderhulst01}, for M106. We note that both surveys are
interferometric, and hence trace only the relatively high ($>10^{19}$
cm$^{-2}$) column density \ion{H}{1}. Extended diffuse \ion{H}{1} in the
outskirts of these galaxies --- where we are most interested --- may be
systematically missed in such surveys.

\section{Analysis Techniques}

Our goal is to measure the spatially-resolved properties of each
galaxy without the need for complete azimuthal averaging. This allows
us to study whether the disk outskirts are azimuthally mixed (as might
arise from radial migration scenarios) or show significant azimuthal
variations (as might occur if outer disks are shaped by recent
accretion events or stochastic star formation). Given the large
angular size of our galaxies, as well as the low pixel-to-pixel noise
in our final mosaic images, we achieve high photometric accuracy
($\sigma_{B-V} \sim 0.1$ mag at $\mu_B=27.5$) over scales of $\sim$ 1
kpc. In analyzing each galaxy, we construct azimuthally distinct
radial surface brightness and color profiles, and decompose the
azimuthal surface brightness variations of the disks into their low
order Fourier modes.  We give details on each method here. All surface
brightnesses are calculated using ``asinh magnitudes''
\citep{lupton99}, which are equivalent to regular magnitudes at high
flux levels, but better behaved at low signal-to-noise.

\subsection{Surface brightness and color profiles}

For each galaxy, we measure the radial surface brightness and color
profiles along six equal-angle radial wedges in the disk plane,
increasing the radial width of each bin with radius in order to
preserve high S/N in the faint outer regions.  By necessity, we use a
constant position angle and ellipticity at each radial bin for these
profiles; the galaxies often show significant variation in isophotal
position angle and ellipticity, and interpretation of the profiles
becomes extremely confused if the isophotal bins are allowed to
wander.  Still, care must be taken in interpreting each profile,
particularly in the disk outskirts where background can begin to mix
with starlight in portions of each wedge due to misalignment of the
aperture with the true, frequently asymmetric isophotes.

We also measure the azimuthally-averaged surface brightness and color
profiles for comparison.  In this case, we do allow the isophotes to
wander, which typically has minimal effect on the qualitative profile
shape.  However, this choice is occasionally non-negligible; for M106,
we found that the surface brightness profiles in both the \emph{B} and
\emph{V} bands show clear Type II (downbending) breaks at 550\arcsec
\ when using varying isophotal parameters, but that this break
disappears completely when using fixed isophotal parameters. Choice of
method thus changes the classification of M106's exponential profile
from Type II to Type I. In M94 and M64, profile breaks appear using
either method, but the changes in slope are much less abrupt in each
case when using fixed parameters over varying parameters.  Using
varying parameters for the azimuthally averaged profile, but fixed
parameters for the angular profiles, also causes the
azimuthally-averaged values not to follow the average of all six
profiles.  In M64, for example, the azimuthally averaged profile
favors the major axis due to changing ellipticity in the outer
isophotes (a property we discuss in more detail in Section~4.2).

In calculating the profiles, we use the median surface brightness of
the pixels in each bin, as it is more robust against contamination
from foreground stars or background galaxies. Comparing these profiles
to those derived from the total flux in each radial bin shows
significant differences in the inner regions, where spiral arms
and \ion{H}{2} regions dominate the light. In the outskirts, however,
the median and luminosity-weighted profiles of the outer regions of
each galaxy are nearly identical, due to our masking of bright sources
in the disk outskirts. While this masking risks excluding light from
bright star forming knots in the disk outskirts, without
high-resolution imaging it is often impossible to differentiate
between background galaxies, foreground stars, and compact sources
within the galaxy itself, and at such faint levels even one such stray
source can dominate the total luminosity of a given annulus or
wedge. This is a known limitation of deep surface photometry
\citep[see, for example,][]{bland05}, and thus to avoid ambiguity with
the background, we choose to measure the properties only of the
diffuse starlight in the outer regions. To mollify the effects of
masking, we compared our images of each galaxy by eye with GALEX FUV
and NUV images to seek out, for example, potential extended star
forming regions, but found that such objects are rare.  Hence, the
populations we sample appear to be representative of the outer disk as
a whole.

Given that we perform surface photometry in the faint outskirts using
our masked and median-binned images, we correct our surface brightness
profiles using sky-subtracted background values measured from these
binned images rather than the unbinned images.  To measure these
values, we place $\sim$50 equal-sized boxes in regions near each
galaxy free of obvious contamination from unmasked sources (in M64's
case, this leaves few regions where we can accurately sample the sky
due the foreground cirrus contamination), and take the median of the
median values of all boxes as the local sky.  The sky uncertainty is
hence the dispersion in the medians, which is quite small
($\sigma_{sky} \sim$0.3 ADU).  Within each box, typical pixel-to-pixel
variance is found to be $\sim$1 ADU, with very little variation
($\sigma \sim$0.1 ADU) from box to box; hence, we subtract the same
sky value from all profiles for a given galaxy.

\subsection{Fourier analysis}

In addition to surface brightness and color profiles, we also conduct
a Fourier mode analysis of the azimuthal surface brightness profiles
of each galaxy, as a function of radius.  This analysis is similar to
that described by \citet{zaritsky97, mihos13a, zaritsky13} and others
to measure lopsidedness in galaxy disks.  We decompose the azimuthal
surface brightness profiles as a function of radius into Fourier
modes:
\[ I(\theta) = \sum\limits_{m}\cos(m\theta+\phi_{m}) \]
where $I$ is the intensity, $m$ is the Fourier mode, $\theta$ is the
azimuth angle, and $\phi_{m}$ is the position angle of the $m^{th}$
Fourier mode.

We measure both the $m=1$ and $m=2$ mode amplitudes in each galaxy,
normalized to the $m=0$ mode (the mean surface brightness in the
annulus), as a function of radius, again using annuli with constant
position angle and ellipticity. Typically, $m=1$ modes are indicative
of galaxy lopsidedness, while $m=2$ and higher modes are related to
repeating patterns such as bars or spiral arms. As such, a measurement
of $m=1$ power in the outer disk can be an indication of a tidal
disturbance that has not had time to settle \citep[but
  see][]{zaritsky13}, while a measurement of $m=2$ power in the outer
disk might indicate extended spiral patterns.  However, $m=2$ modes
may also arise from misalignments between the photometric aperture and
the true isophotal shape, due to asymmetries such as warps or tidal
distortions in the disk.  As such, visual inspection is necessary in
interpreting this type of modal analysis to avoid drawing false
conclusions.

\section{Individual galaxies}

\begin{deluxetable}{l c c c c}
\tabletypesize{\scriptsize}
\tablecaption{Galaxy Properties\label{tab:gals}}
\tablecolumns{5}
\tablehead{\colhead{} &
           \colhead{} &
           \colhead{M 94} &
           \colhead{M 64} &
           \colhead{M 106} \\
           \colhead{} &
           \colhead{} &
           \colhead{NGC 4736} &
           \colhead{NGC 4826} &
           \colhead{NGC 4258}
}
\startdata
(1) RA              &  [J2000]              & 12:50:53.0              & 12:56:43.6                & 12:18:57.5            \\
(2) Dec             &  [J2000]              & +41:07:14               & +21:40:59                 & +47:18:14             \\          
(3) Type            &                       & (R)SA(r)ab              & (R)SA(rs)ab               & SAB(s)bc               \\
(4) Distance        &  [Mpc]                & 4.2\,\tablenotemark{a}  & 4.7\,\tablenotemark{b}   & 7.6\,\tablenotemark{c}  \\
(5) $M_{B_{T}^{0}}$    &                       & $-$19.4                   & $-$19.5                     & $-$20.9                  \\
(6) $(B-V)_{T}^{0}$  &                       & 0.72                    & 0.71                      & 0.55                    \\
(7) $M_{HI}$         &[$10^{8} M_{\astrosun}$]  & 4.00\,\tablenotemark{d} & 5.48\,\tablenotemark{d}   & 35.9\,\tablenotemark{f} \\
(8) $M_{HI}/L_{B}$    & [$M_{\astrosun}/L_{\astrosun,B}$]  & 0.045                   & 0.056                      & 0.101                  \\
(9) $R_{25}$         & [arcmin]                     & 5.6                     & 5.0                       & 9.3                    \\
(10) $R_{25}$         & [kpc]                                   & 6.8                     & 6.8                       & 20.6                   \\
(11) $W_{50}$         & [km s$^{-1}$]                                & 208.5\,\tablenotemark{d} & 304.0\,\tablenotemark{d} & 381\,\tablenotemark{g} \\
(12) SFR$_{H\alpha}$   & [\sfrone]                       & 0.43\,\tablenotemark{d} & 0.82\,\tablenotemark{d}   & 3.82\,\tablenotemark{e} \\
(13) Scale    & [pc arcsec$^{-1}$]                     & 20.4                    & 22.8                     & 36.8        
\enddata

\tablecomments{
  Rows are: Right ascension and declination (1,2), morphological type (3),
  adopted distance (4), absolute B magnitude (5), B$-$V color (6), \ion{H}{1} 
  mass (7), \ion{H}{1} mass per unit blue luminosity (8),
  \mub$=25$ isophotal radius in arcminutes (9) and kpc (10), \
  \ion{H}{1} line width (11), H$\alpha$ star formation rate (12),
  and  physical scale (13). All values come from the RC3 \citep{devau91},
  except for: (a) \citet{radburn11b}, (b) \citet{jacobs09}, (c) \citet{humphreys13},
  (d) \citet{walter08}, (e) \citet{kennicutt98}, (f) \citet{vanderhulst01}, (g) \citet{tully09}.
}

\end{deluxetable}

Here we present the results of our broadband imaging and surface
photometry of these three galaxies. For reference, we present various
global properties of these galaxies in Table \ref{tab:gals}.

Figure~\ref{fig:m94comp} and subsequent figures show a comparison
between our broadband imaging, GALEX FUV and NUV imaging, and THINGS and
WHISP \ion{H}{1} imaging (see Section 2.3). Our broadband imaging is
shown in the upper-left of the figures, with the intensity scale
rewrapped over three ranges of brightness (\mub$<24.6$,
$24.6<$\mub$<26.5$, and \mub$>26.5$) to highlight different regions. We
show the unbinned, native resolution images inside of the \mub$=26.5$
isophote, and the $9\times9$ binned images outside of this isophote in
order to enhance faint, extended features. In the upper-right of the
figures, we show a \bmv\ pixel-to-pixel color map of our broadband data
(at native resolution only). The colorbars on the righthand sides give
\bmv\ values. UV data from GALEX is shown in the lower-left of the
figures (FUV in blue and NUV in yellow), while 21 cm emission is shown
in the lower-right.

We overlay white ellipses of various semi-major axis length on each
image, to provide a visual scale for the surface brightness and color
profiles shown in Figure~\ref{fig:m94profs} and subsequent figures. Each
ellipse uses the parameters (ellipticity and position angle) of the last
best-fit isophote of the unbinned image, and is labeled in arcseconds.
We also plot two red lines to indicate the major and minor axes of these
ellipses, labeled $0^{\circ}$ and $90^{\circ}$, respectively, with
$0^{\circ}$ marking the position angle of the major axis.

Figure~\ref{fig:m94profs} and subsequent figures show surface
brightness and color profiles of the galaxies, plotted as a function
of semi-major axis length (shown in arcseconds and kpc). The colored
lines in the top-left (\emph{B} band surface brightness) and
bottom-left (\bmv \ color) panels of Figure~\ref{fig:m94profs}
represent profiles measured along the corresponding colored wedges
depicted in the inset schematic (solid lines indicate where the
unbinned mosaic was used, and dashed lines indicate where the
$9\times9$ binned mosaic was used). We overplot the
azimuthally-averaged profiles of each galaxy as well using black
squares (unbinned data) and triangles ($9\times9$ binned
data). Characteristic error bars are also shown in each figure,
dominated by the presence of faint, unmasked background sources.
Because this is correlated scatter, the error in color is much less
than the quadrature sum of errors in surface brightness
\citep[see][]{rudick10}.  We also include the radial FUV and NUV
surface brightness (in AB magnitudes, shifted upward by 2 mags
arcsec$^{-2}$ to avoid stretching the ordinate axis of each graph) for
comparison, measured using the same isophotes as the optical data: FUV
is shown in purple and NUV is shown in gold, plotted only to where the
FUV surface brightness begins to flatten into a constant background
value. All surface brightnesses and colors have been corrected for
foreground extinction using the extinction maps of \citet{schlegel98}
as recalibrated by \citet{schlafly11}; we use the coefficients
measured by \citet{yuan13} to derive the extinction in the two GALEX
passbands.

Additionally, we show Fourier $m=1$ and $m=2$ amplitudes (normalized to
the $m=0$ amplitudes; see Section 3.2) as a function of semi-major axis
length in the righthand panels of Figure~\ref{fig:m94profs} and
subsequent figures. $m=1$ amplitudes are shown in the upper right, with
their corresponding azimuthal angle plotted just below, while $m=2$
amplitudes and angles are shown in the bottom-right. Angles are measured
in the plane of each galaxy; $0^{\circ}$ thus marks the major axis at
the galaxy's position angle, increasing clockwise (shown by the red
lines in Figure~\ref{fig:m94comp} and subsequent figures). Blue symbols
are measured from the \emph{B} band images, and gold are measured from
the \emph{V} band images. Both bands typically show good agreement
except in regions of low signal-to-noise.

\subsection{M94 (NGC 4736)}

\begin{figure*}
  \centering
  \includegraphics[scale=0.6]{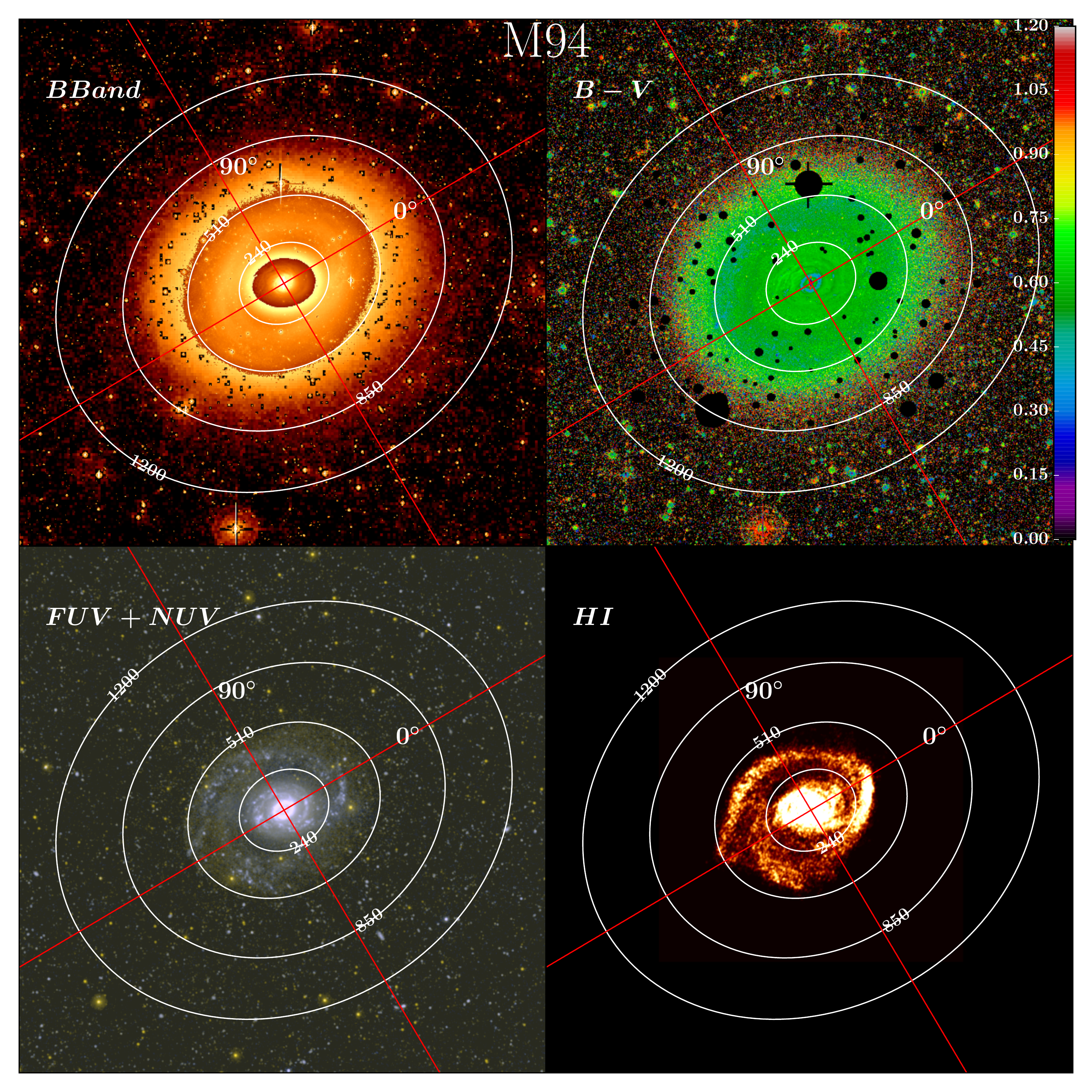} 
  \caption{Images of M94, with ellipses of various semi-major axis
    length overplotted for reference, labeled in arcseconds.  Red
    lines indicate the major and minor axes of the outermost isophote
    of the unbinned image; labeled angles give degrees from the chosen
    galaxy position angle for the Fourier analysis discussed in
    Section 3.2.  From the top left: {\bf 1.)} A subset of our
    \emph{B} band mosaic, rescaled in intensity to highlight ranges of
    surface brightness (\mub$<24.6$, $24.6<$\mub$<26.5$, and
    \mub$>26.5$); the $9\times9$ median binned image is shown outside
    of \mub$>26.5$ to enhance diffuse features; {\bf 2.)}  \bmv
    \ color map; \bmv \ values are shown via the colorbar on the
    righthand side; {\bf 3.)} $FUV$ and $NUV$ false-color image
    constructed from GALEX data, specifically the Nearby Galaxy Survey
    \citep[NGS;][]{bianchi03}; blue denotes FUV data, yellow denotes
    NUV data; {\bf 4.)} \ion{H}{1} image constructed from THINGS data
    \citep{walter08}; 1$\sigma$ rms noise of this image is
    2.6$\times 10^{20}$cm$^{-2}$ \citep{walter08}.  All four plots
    are shown at an identical angular scale.
    \label{fig:m94comp}}
\end{figure*} 

\begin{figure*}
  \centering
  \includegraphics[scale=0.6]{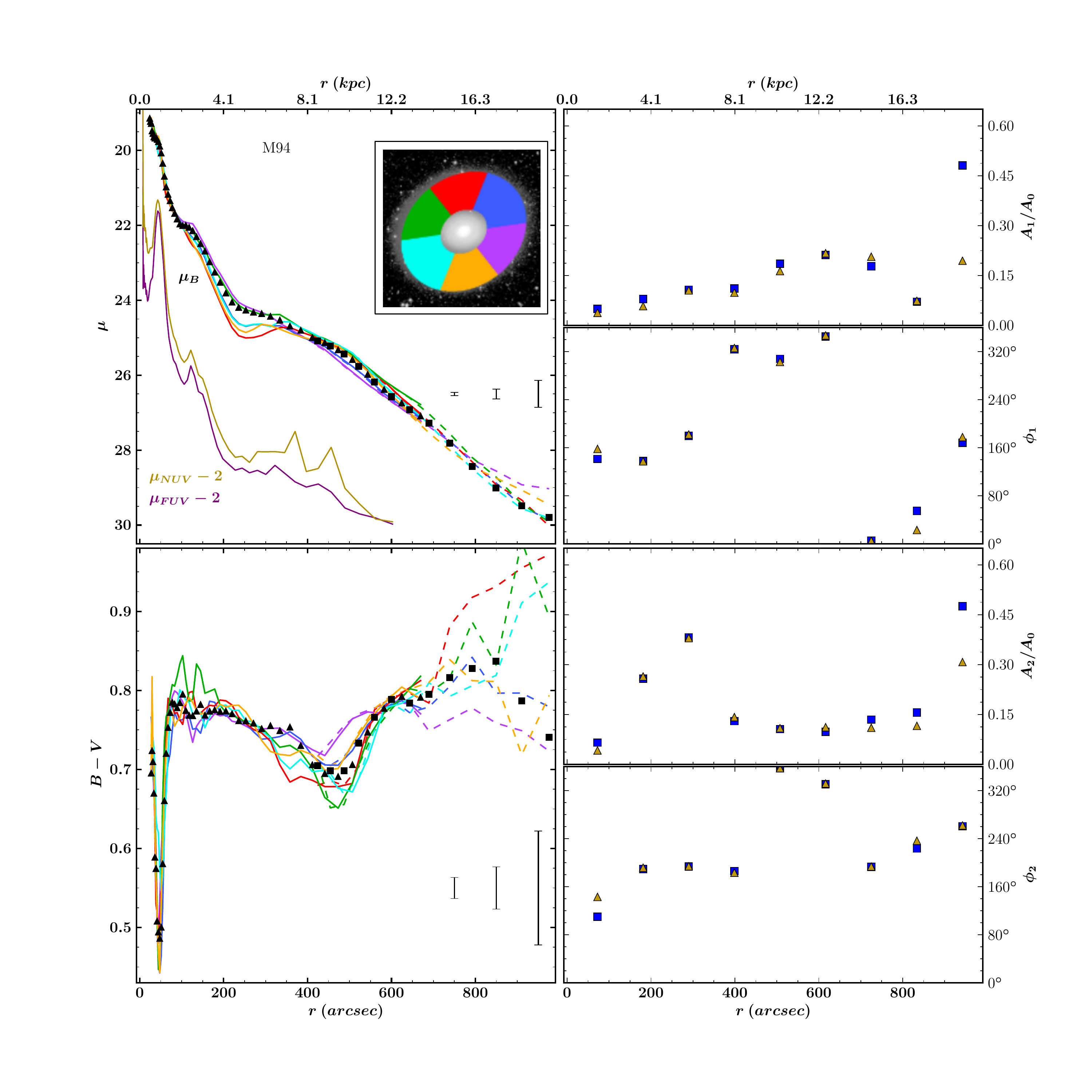}
  \caption{{\bf Top Left:} \emph{B}, \emph{FUV}, and \emph{NUV}
    surface brightness profiles of M94.  Triangles and squares are
    azimuthally-averaged values, while colored lines are profiles
    averaged over the corresponding colored wedges shown in the inset
    schematic.  Triangles and solid lines denote where unbinned data
    is used, and squares and dashed lines denote where the $9\times9$
    binned data is used (see Section 2.2).  The purple and gold lines
    show the FUV and NUV profiles, respectively; UV surface
    brightnesses are shifted brighter by 2 magnitudes to avoid stretching
    the y-axis, and are cut off just before the FUV surface brightness
    profile flattened to the background value.  All values are
    corrected for foreground extinction.  Representative error
      bars are shown for \mub$=$28, 29, and 30.  Error bars are
      smaller than the point sizes for surface brightnesses below
      these values. {\bf Bottom Left:} \bmv
    \ profiles of M94.  Symbols and colors used are the same as the
    top left plot.  UV colors are not plotted.  Values are also
    corrected for foreground extinction. Representative error
      bars are shown for \mub$=$26, 27, and 28. {\bf Top Right:} $m=1$
    amplitude normalized to $m=0$ amplitude (top) and angle (bottom)
    of the azimuthal intensity profile, plotted as a function of
    semi-major axis length (see Section 3.2).  Blue squares indicate
    values obtained from the \emph{B} band image, and gold triangles
    indicate values obtained from the \emph{V} band image.  {\bf
      Bottom Right:} $m=2$ mode normalized amplitude and angle of
    azimuthal intensity profiles, as above.  We assume a distance of
    4.2 Mpc to M94 \citep{radburn11b}, for a disk scale length of
    2.3 kpc (based on the outer disk).
  \label{fig:m94profs}}
\end{figure*} 

M94 (NGC 4736) is part of the Canes Venatici I Cloud
\citep{karachentsev05}, a loose association of galaxies that may be
expanding with the Hubble flow \citep{karachentsev03}. It is hence
fairly isolated: \citet{karachentsev14} list its nearest neighbor as IC
3687, a dwarf galaxy located at roughly the same distance as M94
\citep[$\sim4.5$ Mpc;][]{jacobs09, radburn11b}, but some $\sim3^{\circ}$
($\sim240$ kpc, in projection) away on the sky \citep[see
also:][]{geller83}. The galaxy contains an outer star-forming structure,
often referred to as a `ring', at $\sim200$\arcsec \ (4 kpc), which is
offset in position angle from the bright inner disk. This ring is also
visible in \ion{H}{1} \citep[e.g.][]{bosma77, mulder93}, where it
appears as a set of irregular spiral arms at high column density
\citep[this is also true of its appearance in the UV, e.g. Trujillo et
al. 2009]{walter08}. Despite this unusual morphology, the gas kinematics
show a monotonic rotation curve from the center out \citep{bosma77,
mulder93}, however noncircular motions are prevalent throughout the disk
at small spatial scales \citep{walter08}. There is evidence that these
outer spiral arms, as well as a more strongly star-forming inner ring
inside of 50\arcsec, are the result of Lindblad resonances (Gu et al.
1996, Trujillo et al. 2009; it is also interesting to note that the
inner and outer rings both have approximately the same position angle;
Mulder \& van Driel 1993). Planetary nebulae (PNe) kinematics also show
evidence of flaring in the old stellar populations \citep{herrmann09a,
herrmann09b} that may imply some past perturbation.  Overall, the galaxy
is difficult to characterize, and shows many asymmetric features
indicative of a possible recent interaction, despite its rather isolated
neighborhood.

Figure~\ref{fig:m94comp} shows these asymmetric features, all located
inside of 500\arcsec (10 kpc). The outer spiral arms can be traced in
our \bmv \ color map, the GALEX images, and in the \ion{H}{1}, and show
strong north-south asymmetry and several kinks qualitatively similar to
those found in the grand-design spiral arms of NGC 5194 \citep{dobbs10},
a galaxy known to be interacting with its S0 companion NGC 5195
\citep{toomre72, salo00, durrell03}. Beyond this radius, however, both
the UV and 21cm emission drop off abruptly, leaving only the smooth
optical isophotes. The outer disk profile continues dropping
exponentially with no sign of any break out to at at least 20 kpc
($\sim$ 9 outer disk scale lengths). While the azimuthally averaged
surface brightness profile of the galaxy shows a mild flattening in the
last two points (at \mub$\sim30$), suggestive of transition into a
smooth halo, this is in fact well-modeled by a transition from the disk
to the local background. We also identify a faint plume visible on
the southwest side, the source of an upturn in surface brightness along
the southwestern minor axis wedges (yellow and purple curves) shown in
Figure~\ref{fig:m94profs}; while this plume may be part of M94's disk,
it is too faint (\mub$>29$) to constrain its color.

M94's \bmv\ color profile (bottom left in Figure~\ref{fig:m94profs})
shows a blueward color gradient in the disk between 200\arcsec\ and
500\arcsec\ (4 -- 10 kpc), at which point the gradient reverses
\citep[this behavior can also be seen in the $g - r$ profile shown
by][though we trace the red part of the profile some 200\arcsec \ beyond
the apparent limit of their data]{trujillo09}. The FUV and NUV profiles
both show a spike in surface brightness at the tail end of the blue
gradient, coincident with the outer spiral arms; beyond this point,
however, the UV profiles truncate. This same behavior appears in the
high column-density \ion{H}{1} gas as well, though lower column-density
gas may well exist beyond this radius.

The six angular surface brightness and color profiles of the disk show
little spread beyond this radius as well. Each color profile turns
redward at roughly the same radius, albeit with varying degrees of
sharpness. The southwest side of the galaxy shows the mildest
gradients; from Figure~\ref{fig:m94comp}, this is also where the
spiral arms appear weakest. Otherwise, the average interquartile
spread amongst all six wedges beyond the break in the color profile is
only $\Delta\mu_{B}\sim$0.2 and $\Delta($\bmv$)\sim$0.02.  As such, it
appears that the bulk of the recent star formation in the disk ends
with the outer spiral arms; if star formation occurs beyond this
radius, it is either at extremely low levels ($\mu_{FUV,AB}<5 \times
10^{-5}$\sfrtwo) or very localized such that we smooth over it in our
angular bins as well. However, visual inspection of the FUV image and
the \bmv\ colormap shows no obvious patches of new star formation in
the outer disk.

Additionally, the $m=2$ amplitude weakens outside of this radius,
lending credence to the idea of an azimuthally smooth outer disk. The
strong $m=2$ mode seen inside of 200\arcsec\ (4 kpc) is driven by an
offset between the inner and outer disk, specifically the gap between
the inner disk and the outer spiral arms. We find some evidence of
lopsidedness beyond 300\arcsec\ (6 kpc), though it is mild ($A_{1}/A_{0}
\sim 0.2$). In the final radial bin, the $m=1$ and $m=2$ amplitudes peak
sharply in the \emph{B} band (and to some extent in \emph{V}), however
the azimuthal variations in surface brightness in low S/N regions is
sensitive to background fluctuations, hence caution is warranted in
their interpretation. That said, this final radial bin does encompass
the southwestern plume, which may be driving some of the
non-axisymmetric power.

Given the smoothness of the outer disk, it is worth discussing the
distorted inner disk morphology in more detail. Inside of
$\sim$200\arcsec\ (4 kpc), the disk is offset significantly in
position angle from the outer disk, suggestive of a tilt in
inclination between the two. However, due in part to the noticeable
offset between the \ion{H}{1} kinematical major axis and the optical
major axis of the inner disk \citep{bosma77, kormendy79}, M94 has long
been thought to host an oval distortion, a type of disk instability
similar to a bar but larger in physical scale \citep[a good overview
  of oval distortions can be found in Section 3.2
  of][]{kormendy04}. Early density-wave models suggested that oval
distortions may maintain spiral structure \citep[this was proposed,
  but not explored, by][]{toomre69}, acting in a manner very similar
to bars \citep{kormendy79, athanassoula80, kormendy04}. In
simulations, \citet{trujillo09} found that an oval distortion in the
inner disk provided a good match to M94's observed structure, lending
additional support to the idea.

We examine this idea again using our surface photometry. M94's surface
brightness profile is complex; \citet{trujillo09} stated that M94 may be
considered an antitruncated disk, given the nearly flat profile in the
outer spiral arm region (300\arcsec\ -- 400\arcsec, 6 -- 8 kpc), unless
the inner disk was truly an oval distortion, in which case it would be
better classified as a single-exponential. In contrast,
\citet{herrmann09b} used PNe kinematics to explain this flattening as an
increased importance of a thick disk component in the outer regions. It
is thus interesting to note that the surface brightness profile of the
galaxy shows a larger inner disk scale length (0.8 kpc, measured between
85\arcsec \ and 165\arcsec) along the minor axis (green and purple
curves in Figure~\ref{fig:m94profs}) than the major axis (0.6 kpc). This
difference becomes most notable around 200\arcsec (4 kpc), coincident
with the gap between the inner disk and outer spiral arms. Thus, M94's
surface brightness profile appears more cleanly exponential when the
inner disk-outer-disk gap is avoided.

The smoothness of M94's profile thus provides additional evidence in
favor of a continuous stellar disk, favoring the oval distortion model
over an outside accretion event to explain the galaxy's
structure. This explanation may not contradict the results of
\citet{herrmann09b}; resonances with bars or similar features can
drive vertical heating and lead to larger scale heights \citep[but see
  Minchev et al. 2012]{schonrich09}. While the mild lopsidedness of
the outer isophotes may be evidence of recent interactions,
lopsidedness is extremely common even in isolated galaxies
\citep{zaritsky13}, and may in fact be a signature of misalignment
between the stellar disk and dark matter halo. As such, the idea of
M94 as a solitary, isolated galaxy evolving in an almost purely
secular way appears sound. M94 may thus serve as a particularly
interesting target for future studies investigating the effect of
secular processes such as radial migration on outer disks.

\subsection{M64 (NGC 4826)}

\begin{figure*}
  \centering
  \includegraphics[scale=0.6]{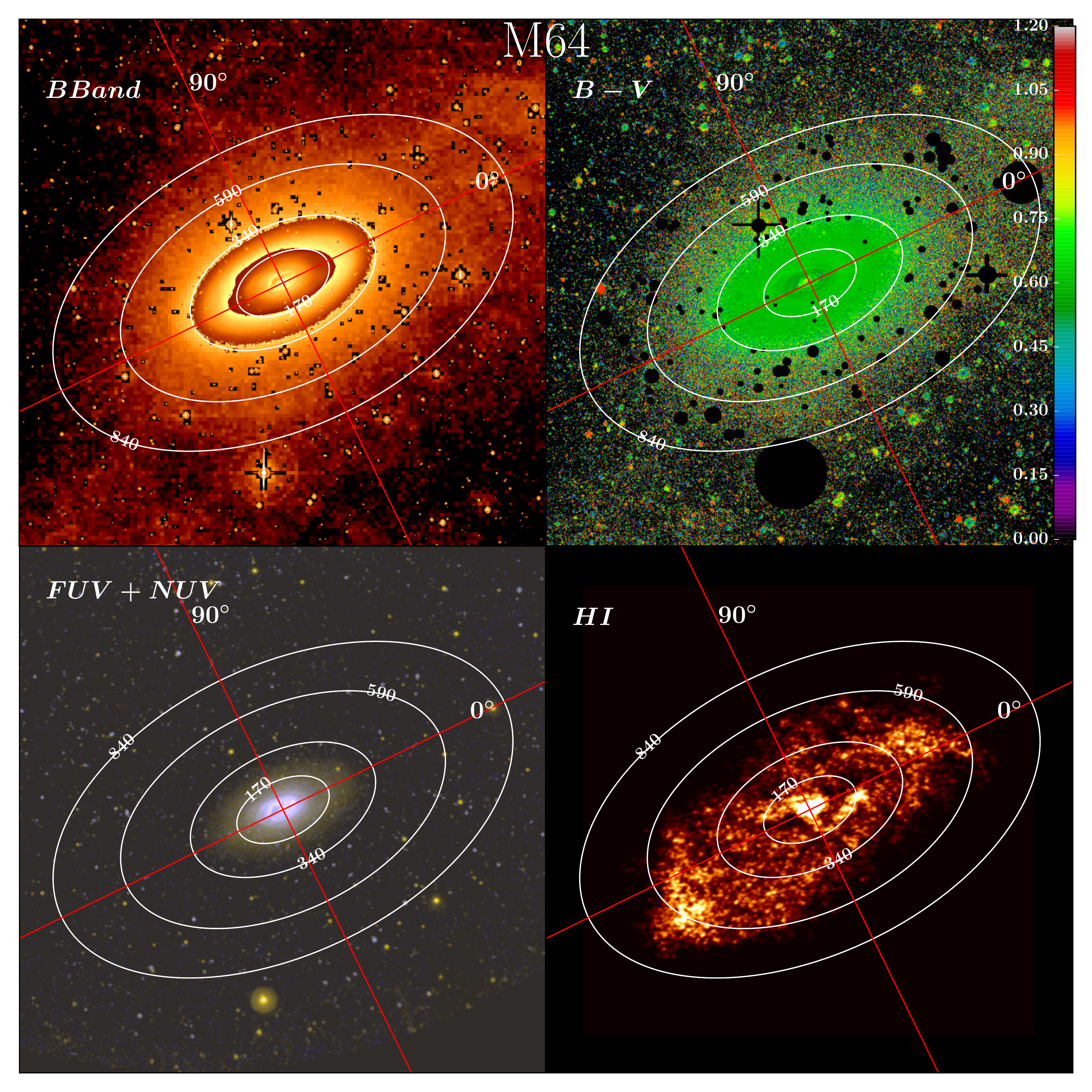}
  \caption{Imaging of M64, using the same layout as Figure~\ref{fig:m94comp}.  
    GALEX data is from the Calibration Imaging survey (CAI);
    M64 appears serendipitously near the edge of the field, and
    exposure times differ between NUV and FUV ($\sim$7000s and
    $\sim$1000s, respectively).  1$\sigma$ rms noise of the
    \ion{H}{1} image is 3.4$\times 10^{20}$cm$^{-2}$ \citep{walter08}.
    \label{fig:m64comp}}
\end{figure*} 

\begin{figure*}
  \centering
  \includegraphics[scale=0.6]{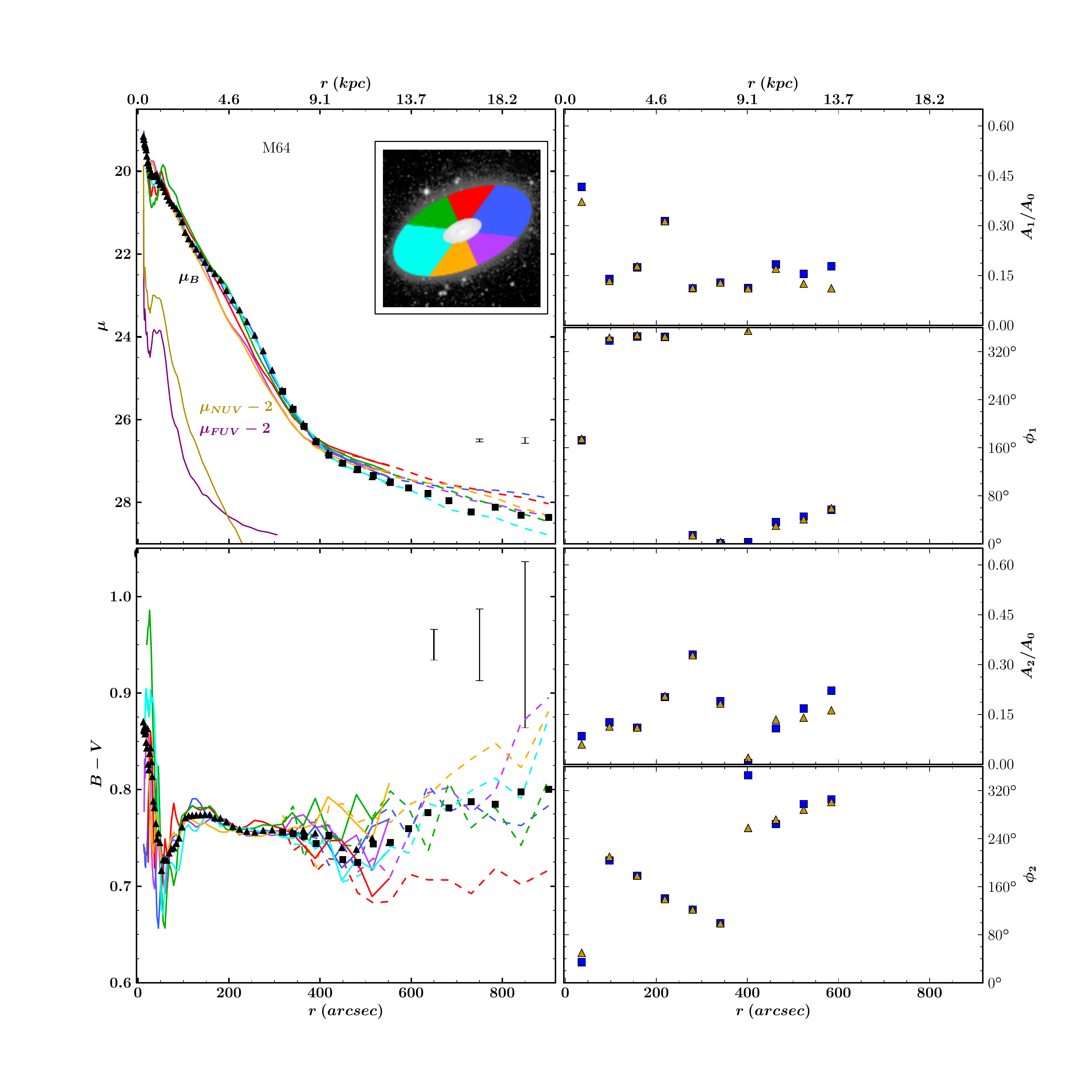}
\caption{Photometric analysis of M64, using the same layout as
  Figure~\ref{fig:m94profs}. The Fourier analysis stops at a smaller
  radius than the photometric profiles due to contamination from
  asymmetrically distributed foreground dust (see text). The limiting
  NUV surface brightness is much lower than the limiting FUV surface
  brightness due to the difference in exposure times between the two
  filters. We show representative surface brightness error bars
    only for \mub$=$27 and 28 for this galaxy, as the limiting surface
    brightness for this galaxy is \mub$\sim$28.5 (see text).  We show
    the same representative color error bars as
    Figure~\ref{fig:m94profs}, however.  We assume a distance of 4.7
  Mpc to M64 \citep{jacobs09}, for a disk scale length of 1.4 kpc
  (based on the inner disk).
\    \label{fig:m64profs}}
\end{figure*} 

M64 (NGC 4826) is a nearby Sa galaxy, alternately referred to as the
``Black Eye'' or ``Evil Eye'' galaxy \citep[else the ``Sleeping
  Beauty'' galaxy;][]{rubin94} due to the prominent dust lane near the
bulge on the northeast side. Interest in the galaxy peaked when
\citet{braun92} discovered that the inner ($<50$\arcsec, 1 kpc) and
outer ($>50$\arcsec) \ion{H}{1} disks counter-rotated with respect to
each other. Subsequent observations by \citet{rix95} revealed that the
entire stellar disk co-rotates with the inner gas disk, implying that
the outer gas disk is an accretion relic.  Detailed study of the
\ion{H}{1} \citep{rubin94} and CO \citep{garcia03} showed evidence of
shocks and radial inflow in the inner disk originating from the
boundary between the two counter-rotating systems.  Indeed,
simulations reproduce such inflows, as angular momentum cancellation
at the boundary between the counter-rotating components leads to gas
infall \citep{quach15}. The galaxy also shows evidence of both leading
and trailing stellar spiral arms \citep{walterbos94}, suggestive of
some disturbance in the stellar disk as well, but nonetheless the disk
shows an extremely regular exponential surface brightness profile out
to $\sim$300\arcsec. Despite the galaxy's counter-rotation and overall
low \ion{H}{1} surface densities, the \ion{H}{1} rotation curve is
quite regular as well \citep{deblok08, walter08}.

We show our imaging of M64 in Figure~\ref{fig:m64comp}, along with the
radial surface brightness, color, and Fourier profiles in 
Figure~\ref{fig:m64profs}. Unfortunately, M64 lies in a region of the sky 
rife with contamination by foreground Galactic dust (or `cirrus'), which
severely limits the depth of our mosaics of this galaxy. Whereas we can
probe surface brightnesses of \mub$\sim 30$ in M94 and M106, here we
begin transitioning into the background at \mub$ \sim28.5$ due to the
surrounding cirrus. Because of the asymmetric nature of the cirrus, we
cannot measure the Fourier modes of the disk beyond 600\arcsec, where
the dust begins to significantly skew the analysis.

Spiral structure is most obviously seen in our \bmv \ colormap, and
appears constrained to within roughly 200\arcsec (4.5 kpc). The disk
color profile is also generally much flatter than M94's. From 
Figure~\ref{fig:m64profs}, we see a mild blueward gradient between 100\arcsec\
and 200\arcsec\ ($\sim$2 -- 4.5 kpc), beyond which the profile flattens
out at \bmv$\sim$0.75. A single mild ($\sim$0.02 magnitude) blueward dip
appears at $\sim$450\arcsec\ (9.5 kpc), and the color trends
continuously redder beyond this radius. The majority of the UV light is
found within the spiral arms; both the FUV and NUV profiles show a steep
decline with radius, reaching $\mu_{NUV}\sim$30 by $\sim$200\arcsec\
(4.5 kpc). Within this radius, the \ion{H}{1} shows weak spiral
structure as well, while at larger radius the gas distribution is quite
irregular and patchy, with extremely low column density \citep[$<
10^{20}{\rm cm}^{-2}$;][]{deblok08}. The rapid decline in UV surface
brightness and very low outer \ion{H}{1} column density both argue that
recent star formation in M64 is constrained to the inner 4.5 kpc.

The presence of both leading and trailing spiral arms in this galaxy
\citep{walterbos94} implies a disturbed stellar disk as well, and
indeed, distorted isophotes are visible just outside of 170\arcsec (4
kpc), and, more weakly, just outside of 340\arcsec (7 kpc) as a slight
protrusion on the galaxy's west side. The angular profiles also show
more scatter beyond $\sim$175\arcsec, with mild but significant power
in the $m=1$ and $m=2$ modes at these radii as well. Close examination
of the images shows that most of the $m=2$ power comes from
misalignment of the elliptical aperture with the galaxy isophotes,
rather than due to any spiral structure. The slow angular slewing of
the $m=2$ mode thus implies a gradual shift in the outer isophotes'
position angle with respect to the photometric aperture.  Indeed, from
our \emph{ellipse} analysis, we find that the position angle steadily
increases by $\sim 20^{\circ}$ between 200\arcsec\ and 700\arcsec (4.5
-- 15 kpc).

The most notable feature we find in M64, however, is the dramatic
antitruncation of the profile beginning around 400\arcsec\ (9 kpc). We
note that the same break is seen in the \emph{R} band profile of
\citep{gutierrez11}. The immediate concern is that this feature is
induced by the foreground cirrus, however a battery of tests
demonstrates that this is not the case. While the cirrus contamination
seems severe in the northwest and southeast sides of our optical
image, its surface brightness in these regions lies around \mub$=28.0$
to $28.5$. As this Type III upbending break begins roughly 2
magnitudes brighter than this level, it cannot be caused simply by a
transition to this background. Additionally, the break can be seen
along every angular cut at the same radius; this would only be true if
the cirrus were evenly distributed around the galaxy, which
Figure~\ref{fig:m64comp} shows is not true. The effect of the cirrus
on the galaxy's surface brightness profile thus appears to be mild;
indeed, it seems strongest only at the largest radii, where the
northwest profiles (blue and red curves) flatten off most quickly, at
the expected background levels (\mub$\sim 28.0$). The eastern major
axis profile (cyan) shows the lowest surface brightness, and is also
the wedge with the least cirrus contamination.

This antitruncation thus appears inherent to the galaxy. Inside of the
break radius (between 200\arcsec \ and 400\arcsec) the interquartile
spread between the six angular profiles is $\Delta\mu_{B}\sim$0.2 and
$\Delta($\bmv$)\sim$0.01, indicating a very uniform stellar population
despite the asymmetry in the disk. Beyond the profile break, all the
angular profiles trend redward, save for the northern (red) wedge. The
reason for the discrepant northern wedge is not evident, even upon
close examination of the image.  It may be related to the cirrus; we
measure a color of \bmv$\sim$0.7 in the relatively bright patch just
to the northwest of M64, which is also the color at which the northern
profile flattens out.  While showing some patchiness, the mean color
of the background near M64 (which includes contributions from cirrus,
unresolved background sources, and residual sky variance) is
\bmv$=$0.85.  Thus, we tested the cirrus' influence on all of the
color profiles using a simple model --- a screen with a uniform
surface brightness of \mub$=28$ and a uniform color of \bmv$=$0.8
overlaid atop a model galaxy with a similar surface brightness and
color profile as M64 --- and found that it begins to affect M64's
color profile at a surface brightness of \mub$\sim$27.0, or a radius
of $\sim$450\arcsec.  As such, it appears that the redward color
gradients beyond $\sim$450\arcsec \ should be attributed to the
foreground cirrus, and not to changes in stellar populations within
the galaxy.  We thus do not consider the color profile beyond
$\sim$450\arcsec \ for the remainder of our analysis.

If the upbending profile seen in Figure~{\ref{fig:m64profs} is due to
a distinct outer disk, it may have been spawned from the interaction
that led to the counter-rotating kinematics seen in M64. If so, its
red colors rule out, for M64, models where anti-truncated outer
disks are built through induced star formation in extended gas
\citep[a mechanism suggested by][]{laine14} and instead favor
scenarios where angular momentum exchange during a merger migrates
stars into the disk outskirts and forms a Type III break
\citep[e.g.,][]{younger07}. However, the profile presents problems
for the latter scenario as well. The models of \citet{younger07}
display mild breaks ($h_{out}/h_{in} = 1.2-1.8$) which occur at
relatively small radius ($R_{b}/h_{in} = 2.5-4$), while the break we
see in M64 happens at much larger radius ($R_{b}/h_{in} = 6$) and is
significantly more dramatic ($h_{out}/h_{in} = 10$). Furthermore,
from our \emph{ellipse} analysis, outside of the break M64's outer
isophotes become much rounder ($b/a=0.7$) than in the inner disk
($b/a=0.5$), which wouldn't arise from simple radial spreading of
the disk. Taken together, these arguments suggest we are not seeing
an outer disk formed through angular momentum transfer during an
accretion event, but instead a profile transitioning from a disk
component to an outer halo.

This alternative interpretation of upbending profiles was also
proposed by \citet{martin14}, who argued that no true Type III disks
exist, and that upbending profiles simply signal the presence of a
stellar halo.  Under the more detailed classification scheme proposed
by \citet{pohlen06}, this would include the Type III-s galaxies
(spheroidal): galaxies with a Type III break that show progressively
rounder isophotes beyond the break radius.  To test this idea further
regarding M64, we fit a variety of models to its surface brightness
profile using \emph{emcee}, a Python-based Markov Chain Monte Carlo
(MCMC) sampling algorithm \citep{foreman13}.  While a double
exponential model provides a good fit to the data, we found nearly
equally good fits for a disk+power law model with a power law slope of
$\alpha \sim$-2, or disk+S\'{e}rsic profiles with indexes $n \sim$0.5
and 4.  Hence in the end we find that the fits do not provide
discrimination amongst the various models.  Given this, the rounding
of the outer isophotes and the poor match to models of outer disk
formation leads us to prefer the disk+halo interpretation for M64's
overall profile.

With all the evidence that M64 has suffered a recent merger, do we see
signatures in its photometric structure? While the galaxy shows some
lopsidedness within $r=$200\arcsec (4.5 kpc;
Figure~\ref{fig:m64profs}), we see no obvious tidal features in the
outer disk that might signal a past accretion event, although the
foreground cirrus precludes us from probing the faintest levels. While
infalling companions can cause significant vertical heating in disks
\citep[e.g.][]{barnes88, toth92, stewart08, kannan15}, it is less
clear how such accretion could have heated M64's disk to form the
smooth, diffuse outer spheroid while also depositing a thin \ion{H}{1}
disk into the rotational plane. This argues that the outer spheroidal
component may predate the merger, rather than forming during the
event. M64 thus appears to be a case of merger-induced quenching,
where the counter-rotating accretion has disrupted the galaxies
gaseous disk and shut off the bulk of star formation throughout the
system, while leaving the stellar disk largely intact.  In this way,
we may be witnessing the end stages of M64's transition into an S0
galaxy \citep[e.g.][]{borlaff14}.

\subsection{M106 (NGC 4258)}

\begin{figure*}
  \centering
  \includegraphics[scale=0.6]{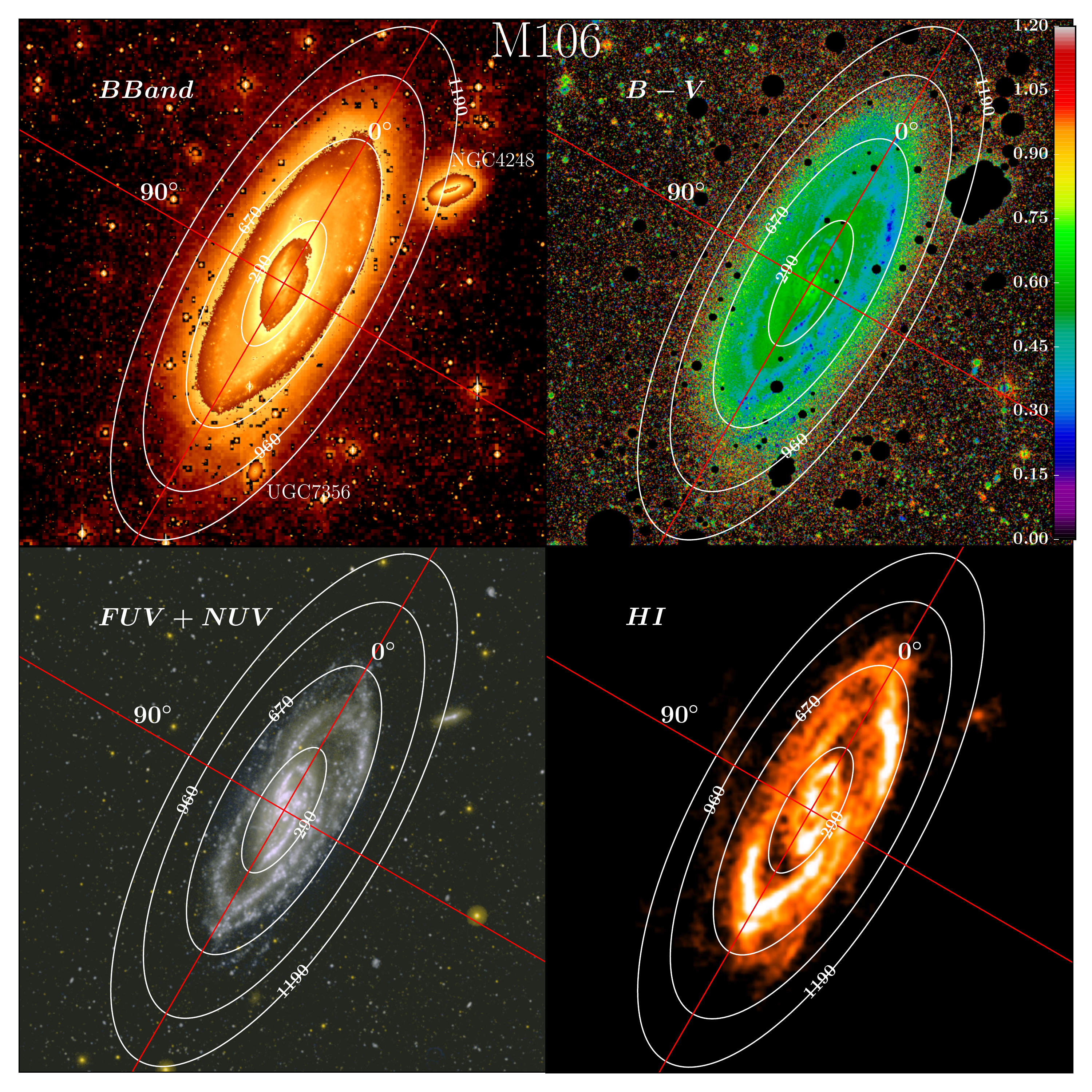}
  \caption{As Figure~\ref{fig:m94comp}, but of M106.  GALEX data is
  from the Deep Imaging Survey (DIS), and \ion{H}{1}
  data is from the WHISP survey \citep[30\arcsec \ resolution
  shown]{vanderhulst01}; 1$\sigma$ rms noise is
  $\sim10^{20}$cm$^{-2}$.  The companion galaxies NGC 4248 and UGC
  7356 are labeled in the \emph{B} band image (upper left).
  \label{fig:m106comp}}
\end{figure*} 

\begin{figure*}
  \centering
  \includegraphics[scale=0.6]{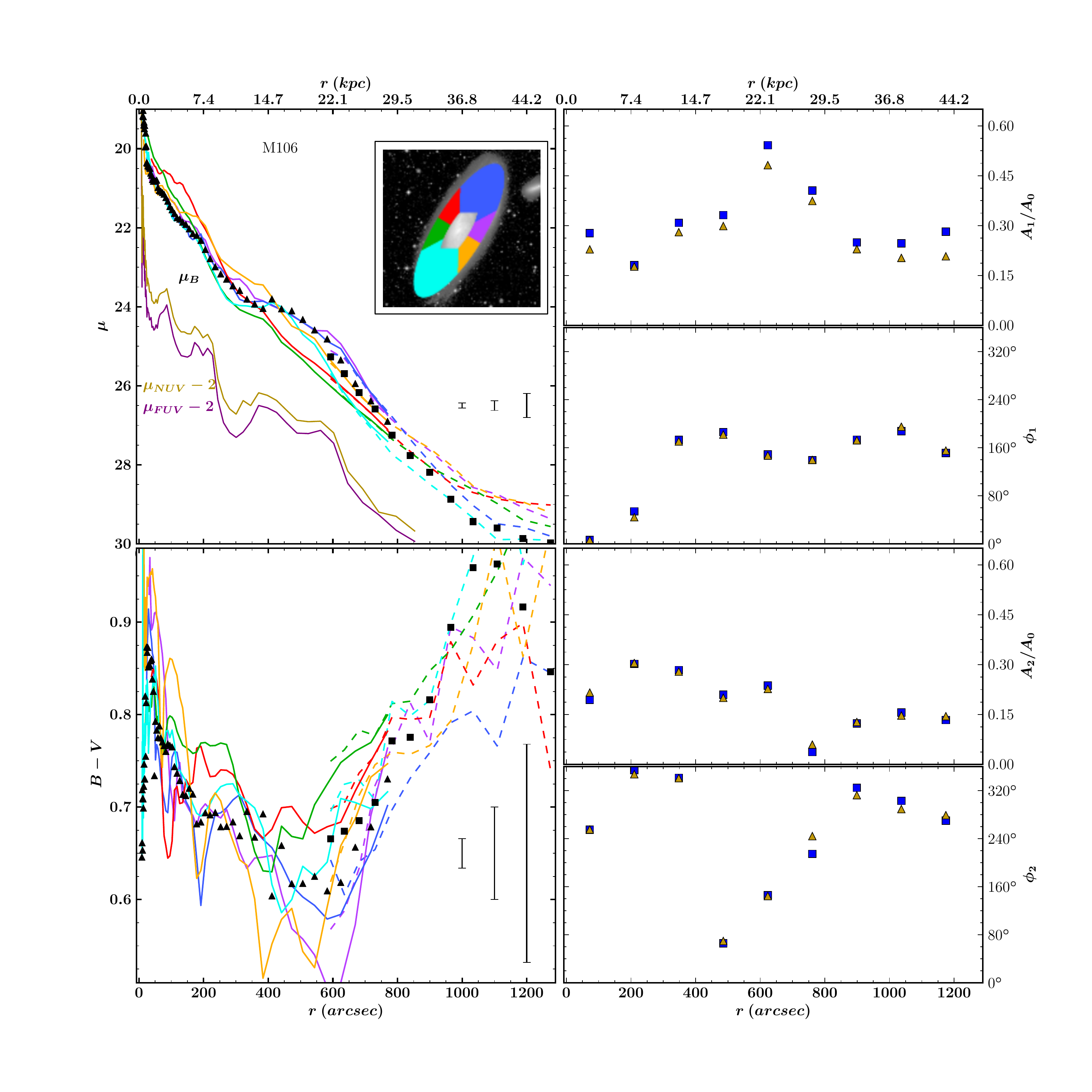}
  \caption{As Figure~\ref{fig:m94profs}, but of M106.  We assume a
  distance of 7.6 Mpc to M106 \citep{humphreys13}, for a disk scale
  length of 6.0 kpc (based on the inner disk).
    \label{fig:m106profs}}
\end{figure*} 

\begin{figure*}
  \centering
  \includegraphics[scale=0.57]{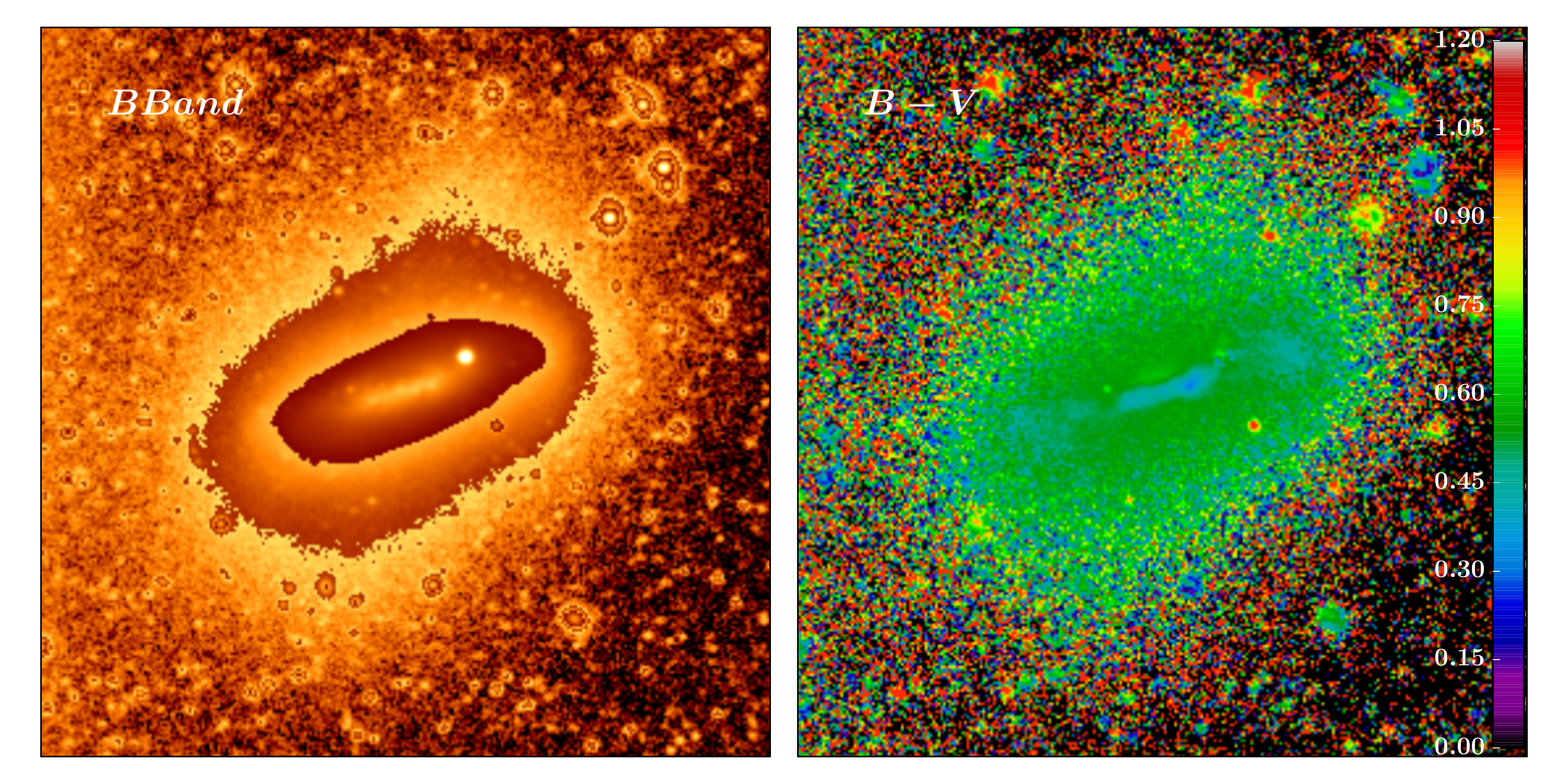}
  \caption{Closeup of NGC 4248.  {\bf Left:} \emph{B} band, shown at
    three levels of brightness (\mub$<23.5$, $23.5<$\mub$<25.5$, and
    \mub$>25.5$) to highlight the warped inner isophotes, the tidal
    features on the galaxy's north and south sides, and the very boxy
    outer isophotes. {\bf Right:} \bmv \ colormap, showing the
    centrally-concentrated star formation.
    \label{fig:4248}}
\end{figure*} 

M106 (NGC 4258) is the brightest member of the Canes Venatici II Group
\citep{fouque92}, and can be considered a Milky Way analogue given its
similar luminosity, Hubble Type, and local environment
\citep[e.g.][]{kim11}. While the central regions of M106 have been
studied extensively \citep[specifically, to determine the origin of a
  pair of offset ``anomalous'' spiral arms;][]{courtes61,
  vanderkruit72, pietsch94}, previous studies on the outer disk are
relatively scarce. M106 is the most massive galaxy of the three
examined in this paper, and is also the only one of the three with
clearly visible satellites. The galaxy NGC 4248, located to the
northwest of M106 (labeled in Figure~\ref{fig:m106comp}), has long
been suspected to be a satellite \citep{vanalbada77}, although
previous studies found no clear evidence of interaction
\citep[e.g.][]{vanderkruit79}. A Tully-Fisher distance places the
galaxy at 7.4 Mpc \citep{karachentsev13}, essentially the same
distance as M106. The galaxy to the southeast, UGC 7356 (also labeled
in Figure~\ref{fig:m106comp}), is also a companion, of much lower mass
\citep{jacobs09, spencer14}. The velocity spread between the three
galaxies is comparable to M106's rotation velocity, suggesting the
companions are bound to M106.  \citet{spencer14} also finds 7
additional probable satellites of low mass ($-12>M_{V}>-17$) within
200 kpc (projected) of M106, indicating a fairly rich local
environment.

We present multiwavelength images of M106 in
Figure~\ref{fig:m106comp}, and the photometric profiles in
Figure~\ref{fig:m106profs}. M106 has more prominent UV emission than
either M94 or M64, reflecting a higher star formation rate (SFR);
indeed, M106's H$\alpha$-derived SFR is 3.82 $M_{\astrosun}$ yr$^{-1}$
\citep{kennicutt98}, compared to 0.43 $M_{\astrosun}$ yr$^{-1}$ and
0.82 $M_{\astrosun}$ yr$^{-1}$ for M94 and M64, respectively
\citep{walter08}.  A rough estimate using the RC3 colors and the \bmv
\ to $M/L$ conversion factors of \citet{bell01} also implies that M106
has a specific star formation rate roughly a factor of three higher
than the other two galaxies. \citet{thilker07} classified M106 as a
Type 1 XUV disk, defined as a galaxy containing highly structured UV
complexes beyond where the UV-derived star formation rate surface
density drops below $\Sigma_{SFR}=3 \times 10^{-4}$~\sfrtwo.  Again,
our \bmv\ colors trace the UV emission well; of note are the extended
spiral arms beyond 300\arcsec\ (10 kpc), and the plume of UV emission
and \ion{H}{1} on the galaxy's south side (near the companion UGC
7356).  This plume, and another on the galaxy's north side (including
some of the diffuse UV light inside of 600\arcsec), drive the
XUV disk classification.  Three diffuse patches of UV emission can
also be found tracing the extremely faint tail of \ion{H}{1} extending
from the galaxy's east side (seen crossing the 965\arcsec \ ellipse in
Figure~\ref{fig:m106comp}), and are also visible as small patches of
diffuse light in our optical imaging (we note that these are not
visible in Figure~\ref{fig:m106comp}, as each patch is only a few
arcseconds in radius and quite faint).

We see a sharp decline in the UV emission beyond the XUV disk, at
roughly 600\arcsec\ (22 kpc), yet the optical light continues to show
an exponential profile well beyond this radius. From
Figure~\ref{fig:m106profs}, the \bmv\ color profile turns sharply
redward at this radius as well. As with M94 (and to some extent M64),
all angular cuts show the same behavior; the spread amongst the cuts,
however, is much greater than in M94 or M64: $\Delta\mu_{B}\sim$0.5
and $\Delta($\bmv$)\sim$0.07. This is due to irregular isophotes; the
minima in the color profiles occur at increasingly large radius from
the northeast (red) curve to the west (purple) curve. This trend
follows the morphology of the spiral arms, which are stronger on the
east side of the galaxy. The northmost (blue) profile breaks the
pattern, however it is clear from Figure~\ref{fig:m106comp} that the
spiral structure is much weaker along this cut. These red disk
outskirts extend to at least 40 kpc, some 20 kpc beyond the apparent
UV cutoff radius in this XUV galaxy and again appears devoid of strong
spiral structure: note the decline in the $m=2$ amplitude beyond
$\sim$600\arcsec\ (22 kpc), as seen in Figure~\ref{fig:m106profs}.

The two bright dwarf companions present one very clear difference
between M106 and the other two spirals in our study. Our deep imaging
shows clear evidence of tidal distortion in the brighter companion NGC
4248 (Figure~\ref{fig:4248}): its low surface brightness outer
isophotes are extremely boxy, and offset in position angle from the
dwarf's inner regions by almost 45$^{\circ}$. In this aspect, it
appears quite similar to NGC 205, a satellite companion to M31 of
comparable luminosity.  Like NGC 4248, NGC 205 also shows recent star
formation very near its center \citep{cappellari99} along with
isophotal twisting \citep{choi02} and a strongly boxy morphology in
its outer regions as well \citep[seen in an image taken by S. van den
  Bergh, presented in][]{kormendy82}. A study of RGB star kinematics
in NGC 205's outskirts revealed strong high-velocity tails and a
reversal in the direction of rotation beyond 1 kpc, indicative of
tidal interactions with M31 \citep{geha06}. The boxiness in NGC 4248's
outer isophotes appears to result from an overlap of an extended
elliptical component with symmetric tidal features extending from the
north and south sides of the galaxy. Though NGC 4248 is tentatively
classified as irregular \citep{devau91}, its warped inner isophotes
(Figure~\ref{fig:4248}) and rotating \ion{H}{1} \citep{vanalbada77}
imply a disklike structure. This galaxy may thus serve as an example
of a low mass disk being tidally transformed into a dwarf elliptical
more akin to NGC 205, an intriguing idea in the context of M106 as a
Milky Way (or M31) analog.

In turn, the effect of the companions on M106's outer disk may be seen
in the disk's visible warp, seen both in the \ion{H}{1} kinematics
\citep{vaneymeren11} and in the generally high $m=1$ amplitude at all
radii in this galaxy (Figure~\ref{fig:m106profs}). The $m=1$ mode
peaks around 600\arcsec, near the outer spiral arm radius; this peak
is due to the southern plume near UGC 7356. Plumes of diffuse
starlight and \ion{H}{1} are also present on the galaxy's north and
south sides along the major axis; the west half of the southernmost
plume shows fingers of UV light and blue colors indicating induced
star formation, and its proximity to UGC 7356 is suggestive of tidal
interaction with that companion \citep[although proximity to tidal
  features is not an unambiguous indication that a satellite has
  generated the disturbance; a clear counter-example is the southern
  tail of M51, generated by the interaction with its companion on the
  galaxy's north side;][]{rots90, salo00}.

As stated in Section 3.1, M106's classification as either a Type II
(downbending) or Type I (unbroken) disk changes depending on the
method used to measure its surface brightness profile. This can be
seen in the behavior of the angular cuts in
Figure~\ref{fig:m106profs}, in that the eastern side of the galaxy
(red and green profiles), where the spiral arms appear weakest, show a
profile contiguous with the outer disk, while the spiral arms induce
an excess of light over this profile in all of the other angular cuts
that appears as a Type II break. This behavior hearkens to the study
by \citet{laine14}, who found that Type II disk breaks tend to follow
morphological features such as spiral arms, lenses, or rings. The
severity of the break thus depends on how closely such features are
followed in the isophotal analysis; in some cases the choice of
isophotes can mask the presence of a disk break entirely or introduce
one where none exists. These effects are most obvious in bluer
wavelengths due to the presence of high-mass stars in the star-forming
regions along spiral arms. In the near-infrared \citep[a fairly robust
  tracer of stellar mass;][]{sheth10}, \citet{laine14} note similarity
between the inner disk scale lengths of Type I and Type III
(antitruncated) disks and the outer disk scale length in Type II
disks.  Given that M106's profile appears unbroken absent the presence
of any spiral arms, one might postulate that the term ``break'' is a
misnomer, and that the inner disks of Type II galaxies are actually
elevated in surface brightness over the baseline outer disk due to
e.g. recent star formation. This would imply that the differences
between Type I and II disks are purely morphological, rather than the
product of different formation histories; Type III disks would thus be
the true outliers, which is consistent with their comparitive rarity
\citep[only $\sim$20\%--30\% of disks show Type III
  breaks;][]{erwin08, laine14}.

Examining the very outermost regions of M106's disk, we again see a
leveling-off of the azimuthally-averaged surface brightness profile in
the last few data points, however as with M94 it appears to be
well-modeled simply by a transition into the local background. However,
we note that the profiles that follow the galaxy's minor axis (red,
green, purple, and yellow) are elevated in surface brightness above the
major axis at this extended radius. The patchy diffuse light that can be
seen outside of 960\arcsec\ in Figure~\ref{fig:m106comp} likely accounts
for this behavior. Given the asymmetry in M106's isophotes, it is
unclear if this patchy light is part of M106's outer disk or instead
represents an inner halo or thick disk; regardless, we can confidently
state that M106's disk extends to at least 1100\arcsec\ (40 kpc, or
$\sim$6.5 disk scale lengths), making it nearly twice as large in
physical size as either M94 or M64.

The tidal features in this galaxy --- the southern and northern
optical plumes and weak \ion{H}{1} tails --- are most likely to have
originated through tidal interactions with its nearby, bound
companions, rather than through a flyby interaction with a more
massive companion. A stronger encounter would likely induce more
dramatic tidal response, but we see no evidence of elongated tidal
tails in M106's vicinity to a limiting surface brightness of
\mub$=29.5$.  The nearest bright galaxy to M106 is NGC 4144
\citep[$M_{B}\sim -18$, assuming a distance of 7.5 Mpc;][but see
  Jacobs et al. 2009]{devau91, seth05}, located some 240 kpc from M106
on the sky \citep{karachentsev14}. Given this separation, and NGC
4144's relatively low luminosity, M106's nearby companions certainly
have the strongest tidal influence, and NGC 4248's boxy outer
isophotes confirm that it is tidally interacting with M106 at some
level.

While strong encounters tend to drive centrally concentrated starburst
activity \citep[e.g.][]{barnes91, hernquist92, hernquist95, cox08,
  hopkins09, powell13, moreno15}, weaker tidal interactions with
satellite galaxies may incite a less-dramatic but longer lived
response in the disk outskirts as they orbit the primary over longer
timescales. While these low mass interactions may be less efficient at
inducing star formation throughout the host \citep{cox08}, even a weak
starburst in the outer disk may transform the structure and stellar
populations of these low surface brightness regions. It may thus be
interesting to consider the possibility that NGC 4248 (and to a lesser
extent UGC 7356) may be shepherding the gas in M106 in such a way as
to produce these outer spiral arms and, at least potentially, trigger
star formation in the otherwise low-density outer \ion{H}{1}. However,
compared to the total extent of the disk, the star formation in M106
is {\it not} greatly extended; while \ion{H}{1} is present at large
radius \citep{wolfinger13}, only within 20 kpc ($\sim$3 disk scale
lengths) is the gas dense enough to form stars. This stands in
contrast to the case in the nearby face-on spiral M101, where
interactions with its nearby companions have triggered star formation
in the galaxy's diffuse outer disk \citep{waller97, mihos13a}. Why,
then, was star formation triggered in the outskirts of M101, but
seemingly not in M106?

The answer may lie in the fact that in addition to driving tidal
resonances in galaxy disks, interactions can also drive non-planar
responses including warps and disk heating, which have the potential to
shut down star formation. The relative efficacy of these different
processes depends not only on the mass ratio of the encounter but also
the orbital properties of the encounter. While M101 has a single close
satellite (NGC 5477), the galaxy's marked asymmetry and its \ion{H}{1}
kinematics both argue for a single prograde encounter with the more
massive and distant companion galaxy NGC 5474 \citep{mihos12, mihos13a}. In
contrast, M106 has {\it two} close companions, one of which (NGC 4248)
is more massive than M101's close satellite NGC 5477. If the orbital
geometry of these satellites is highly non-planar, the two working in
concert may tip the dynamical balance towards disk heating rather than
tidal compression, suppressing star formation in the outer disk. M106
thus may be an interesting test case concering the influence of fairly
massive dwarf satellite galaxies on the star-forming properties of the
host, which may be of particular interest in Milky Way studies given the
presence of the Magellanic Clouds.

\section{Discussion}

Despite different local environmental conditions and interaction
histories, we see consistent behavior in the photometric properties of
these three galaxies' outer disks.  In M106 and M94, the onsets of
redward gradients in their color profiles correspond to truncations in
the UV surface brightness and 21cm emission tracing high
column-density \ion{H}{1} gas.  In M64, the UV emission is constrained
to the central disk, and the colors flatten beyond the UV truncation
to a similarly red color.  The high column-density \ion{H}{1} is more
extended in this galaxy than in the other two, but is globally at much
lower density and hence non-star-forming.  What is consistent across
all three galaxies is a lack of strong azimuthal color variation in
the outer disks, with the interquartile spread in color beyond the
break radius always $<0.1$ magnitudes (and significantly less in the
cases of M94 and M64).

Spiral features also seem to vanish beyond the UV truncation radius. All
three galaxies show only mild evidence of azimuthal asymmetry in their
outer isophotes, the strongest present in M106, with no evidence of
faint extended spiral structure. In these three galaxies, at least, this
appears to imply a natural division between the ``inner'' and ``outer''
disks; ``outer'' disks may be defined as the region beyond any evident
spiral features and devoid of new star formation, yet still following an
exponential surface brightness profile. Here we address the constraints
placed by our deep surface photometry on the stellar populations in
these outer disks, and compare to studies of outer disk populations in
other galaxies. We also consider the role local environment plays in
shaping each galaxy's outer disk; in tandem with the inferred stellar
populations, these constraints can provide useful clues to the formation
and evolutionary histories of outer disks.

\subsection{Outer disk stellar populations}

The similarity in outer disk colors for each galaxies implies a
similarity in stellar populations. In all three galaxies, the outer
disks display \bmv colors of approximately 0.75--0.8 at a surface
brightness of \mub $\sim$ 27.5. These colors appear robust against the
color uncertainty, which is dominated by fluctuation in the background
of the order $\sigma_{B-V} \sim\pm 0.1$ mags at these surface
brightnesses (see Section 2.2). These colors also appear independent
of the mean background color (as introduced by faint cirrus,
  unresolved background sources, and residual sky variance); while
the background near M64 is fairly red (\bmv = 0.85), near both M106
and M94 it is significantly bluer, \bmv=0.4-0.5. Indeed, we find
similarly red colors in the outskirts of three other disk galaxies we
recently studied\footnote{We note, however, that in the outer disk of
  M101, we find significantly bluer colors
  \citep[\bmv=0.3--0.5;][]{mihos13a}, in congruence with the galaxy's
  classification as an XUV disk \citep{thilker07}. This argues that
  red outer disk colors are not a universal and systematic artefact of
  our instrumental setup, such as the extended wings of the PSF. See
  the Appendix for more details.} --- M96, M95 \citep{watkins14}, and
M51 \citep{watkins15} --- making this color of \bmv = 0.8 a natural
anchor-point from which to study the outer disk populations of our
galaxies. While broadband colors suffer from the well-known
age-metallicity degeneracy \citep{worthey94}, these colors can still
place some constraints on the outer disk stellar populations. As a
fiducial reference, \bmv=0.8 is a typical integrated color of an S0a
type galaxy \citep{roberts94}, implying a fairly evolved population.

To explore population constraints in more detail, we model the
integrated colors of stellar populations built via a variety of star
formation histories and metallicities, constructed using the software
SMpy, a Python-based SED modeling code based on the \citet{bruzual03}
population synthesis models \citep[described in:][]{duncan15}. We
constrain these models using the surface brightness and color of the
outer disks, as well as the upper limits on their inferred star
formation rates (SFRs). At the radius where the disk colors reach
\bmv=0.8, we do not detect significant FUV flux in any of the
galaxies; in M106 and M94 this places a limit on the SFR of
($\lesssim$3-5$\times 10^{-5}$\sfrtwo). While the limit is higher for
M64 due to the FUV image's short exposure time ($\sim
10^{-4}$\sfrtwo), it is not so high as to significantly alter our
conclusions for this galaxy.

Applying these constraints, we run models using both
exponentially-declining histories -- ${\rm SFR(t)} \propto
e^{-t/\tau}$ -- as well as delayed exponential histories -- ${\rm
  SFR(t)} \propto te^{-t/\tau}$ \citep{lee10, schaerer13}. We adopt
varying decay rates ($\tau$) and metallicities for a 10 Gyr timespan,
assuming a \citet{chabrier03} IMF. A constant star formation history
is ruled out by the low current SFR; the timespan required to build
enough stars to match the total \emph{V} band luminosity within the
\mub$=27.5$ annulus in each galaxy is $>20$ Gyr. Between exponential
and delayed exponential histories, consistent behavior emerged: for
solar metallicity and below, current-day colors become too blue if
$\tau \gtrsim 2$ Gyr, signifying a stellar population dominated by old
stars. Metallicities below [Fe/H] $\sim -0.7$ are ruled out, as these
populations produce colors which are too blue regardless of the choice
of $\tau$.

While these colors suggest old and only moderately metal poor ([Fe/H} $>
-0.7$) populations, significant ambiguity remains due to the
age-metallicity degeneracy inherent in broad-band colors. How then do
these results compare to other studies of the outskirts of nearby disks
using resolved stars, which more directly probe the ages and
metallicities of stellar populations? 

Resolved imaging studies show a variety of stellar populations present
in the outskirts of disk galaxies, indicative of diverse star forming
histories.  For example, the outskirts of NGC 300 and NGC 7793 are
populated almost entirely by RGB stars \citep{vlajic09, vlajic11},
while a sizable AGB population was found in the outskirts of NGC 2403
and M33 \citep{davidge03, barker07}.  M83, a galaxy known to have
highly extended star formation \citep{thilker05, bigiel10}, contains RGB,
AGB, and red supergiant (RSG) stars in its outskirts
\citep{davidge10}.

The inferred metallicities of resolved outer disk populations also
show significant variation from galaxy to galaxy. In NGC 300,
\citet{vlajic09} found evidence for a metallicity gradient in the
outer disk, with metallicities spanning the range [Fe/H]$ = -0.5
{\rm\ to} -1.0$.  Similarly low metallicities ([Fe/H]$\sim -1.0$) have
been discovered in the resolved outer disk populations of NGC 2403 and
M33 \citep{davidge03, barker07}, and even lower metallicites are
inferred in NGC 7793's outer disk \citep[\feh$\sim
  -1.5$][]{vlajic11}. Such low metallicities, if characteristic of
outer disk populations in general, would lead to colors much bluer
than we find in M106, M94, and M64, even taking into account age
effects.

However, those studies focused on fairly low mass systems, smaller in
mass than the three galaxies in this study \citep[estimated from their
  maximum rotation velocities listed in the HYPERLEDA
  catalogue;][]{makarov14}, though NGC 2403 is very near in mass to
M94 and M64. If we assume that outer disk populations follow their
host galaxies' behavior on the well-known galaxy mass-metallicity
relationship \citep[e.g.][]{tremonti04}, these galaxies would have a
higher mean metallicity in their outskirts. Indeed, higher
metallicities are inferred in the outer disk populations of the bright
spirals M31 ([M/H]$\sim$-0.3 -- -0.5; Worthey et al. 2005, Gregersen
et al. 2015), M81 ([M/H]$\sim$-0.4 -- -0.7; Williams et al. 2009), and
M83 \citep[metallicities ranging from $\sim$20\% solar to nearly
  solar;][]{davidge10}. At these metallicities, the integrated colors
of the disk would be significantly redder, in line with our deep
surface photometry presented here. We also note that metal-rich
populations in resolved star studies can be systematically missed due
to the the faintness of the metal-rich RGB \citep[e.g.][]{rejkuba05,
  harris07}, complicating comparisons between those studies and deep
surface photometry. It may thus be of interest to do more studies
directly comparing integrated light colors with resolved photometry in
order to better constrain the biases inherent in both methods.  We
discuss one such bias, the galaxy PSF's influence on our measured
colors in the outer isophotes, in the Appendix, though we believe it
to be small within \mub$<$27.5.

The uniformly red colors, azimuthally smooth distribution, and
inferred old, moderately (but not extremely) metal-poor stellar
populations at large radius in these galaxies thus place constraints
on the formation history of their outer disks. Disk building via
continual low-level star formation in the outer disk appears ruled
out: such a model would lead to much bluer colors than we observe, and
the current rate of star formation is too low to build the amount of
light we see in the disk outskirts in a Hubble time.  Instead, radial
migration \citep{roskar08} emerges as the most likely candidate for
disk building at large radius, given the red stellar populations in
all three galaxy outskirts, as well as the U-shaped color profiles in
the two galaxies still actively forming stars in their inner
regions. That said, it is unclear just how far out radial migration
can drive stellar populations. Radial migration requires the presence
of non-axisymmetric structure such as bars or spiral arms
\citep{sellwood02}; significant migration into the outer disk would
require the same mechanisms \citep{minchev12, roskar12}.  In these
three galaxies, we find stars extending out to 3--4 scalelengths
beyond the edge of the spiral arms; if this behavior is common in
other galaxies, it may present a challenge for disk migration models
as well. While additional spreading of the outer disk may arise from
transient, tidally-driven outer spirals, the galaxies studied here
live in fairly low density environments and display no such tidal
features. We therefore look forward to new dynamical modeling of disk
galaxies that will examine these issues in more detail.

\subsection{Environmental influences}

Under the hierarchical accretion paradigm, galaxy disks are built
continually over time, from the inside out, as material from the
surrounding environment (both baryonic and not) continually bombard
the disk. This accretion can grow disks by depositing stars in disk
outskirts \citep{stewart09}, triggering extended disk star formation
\citep{whitmore95, weilbacher00, smith08, powell13}, or moving stars
outwards through tidal heating or radial migration \citep{roskar08,
  koribalski09, khoperskov15}. Yet regardless of how outer disks are
built, one would expect to see signatures of this process in these
very faint, highly extended regions, where dynamical times are long
and material is more loosely bound. However, in the three galaxies
studied here, evidence for such accretion signatures is lacking. In
M94, the most isolated galaxy of the three, we see only mild
lopsidedness in its isophotes, and one extremely faint plume at the
very outer edge of the disk, implying a much less chaotic formation
history. While M64's past interaction history seems to have greatly
damaged the gaseous disk, driving \ion{H}{1} inwards and shutting off
disk-wide star formation, we again see no evidence for discrete tidal
streams. Finally, in the case of M106, a large and luminous disk
galaxy with two known and many more suspected satellites, the tidal
signatures we do observe are rather weak and likely driven by the two
luminous satellites. While we see no strong tidal features in any of
the galaxies studied here, the similarity in the star forming
properties and stellar populations of their outer disks raises the
question of what impact, if any, the local environments might play in
shaping their outer disks.

On large scales, the environments of the three galaxies are similiarly
devoid of massive companions. Within 1 Mpc, M94 has only a handful of
neighbors, all significantly lower in luminosity \citep[the most
  luminous being NGC 4365, with $M_V \sim -18$;][]{devau91,
  jacobs09}. M64 may be even more isolated than M94; its brightest
neighbor is the dwarf NGC 4789A, with $M_{V} \sim -14$
\citep{devau91,jacobs09}. M106 resides in a somewhat richer
environment, with several modestly bright companions within 1 Mpc,
although none approach M106 in luminosity. M94 and M64 thus might be
considered extremely isolated, while M106 resides in a moderately
denser but still fairly sparse environment, more similar to the Local
Group (though with no massive companion analogous to M31).

On smaller scales, however, the local environments of the galaxies do
appear different. Looking for satellite galaxies with $\sim$100 kpc, a
scale comparable to the Milky Way's satellite system, neither M94 nor
M64 have luminous satellites, while M106 has the two mentioned
previously: NGC 4248 and UGC 7356. We thus find three levels of
environmental influence amongst these three galaxies: M94, being very
isolated and apparently undisturbed (Figure~\ref{fig:m94comp}), may be
evolving purely secularly; M64, while also very isolated, likely
suffered a recent merger that greatly affected its own morphology and
star forming properties (Section 4.2); and M106, living in a denser
environment, is presumably being influenced most by its dwarf
companions.

Despite their different local environment, M94 and M106 both have
similar outer disk structure: a set of extended, star-forming outer
spiral arms, beyond which the disk is smoothly distributed and
contains an old stellar population. If M94's outer spiral arms are
formed secularly, and if M106's outer spiral arms are formed via weak
interactions, this implies two very different paths toward a
qualitatively similar result. We see no sign of a strongly perturbed
outer disk in M106, despite the presence of its satellites. This
contrasts with the case of M81, which is similar in luminosity and SFR
to M106 \citep{kennicutt98}, but where its two more massive companions
M82 and NGC 3077 have disrupted its disk outskirts
\citep{vanderhulst79, yun94, okamoto15}.  The effect of the satellites
on M106's disk appears much gentler --- the galaxy may lie in
something of a sweet spot, with companions massive enough drive spiral
structure \citep{weinberg95, oh08, choi15} and mediate radial
migration to build the outer disk, but not massive enough to
significantly disrupt it once formed.

The situation for M64 is somewhat more muddled. At first glance, the
presence of a Type III upbending break in a post-merger galaxy is
consistent with the idea that Type III breaks are driven by strong
interactions \citep{laine14} or accretion events \citep{younger07}.
However, as argued in Section~4.2, the properties of M64's outer
component are a poor match for either the induced star formation model
or the angular momentum transfer model \citep{younger07}.  Instead,
the changing photometric profile is better explained as a disk-halo
transition. That said, the halo is relatively bright: with \mub $\sim$
27 at 10 kpc, it is significantly higher in surface brightness than
that of the Milky Way or M31 \citep{morrison93, gilbert12}.

If this outer profile is indeed a simply a stellar halo, then in M64
we are seeing a smooth and largely unbroken Type I exponential disk
extending all the way out to where it becomes lost in the halo light,
at 6 disk scale lengths. The disk is red and azimuthally smooth, save
for the very inner regions where some residual star formation
continues. With star formation otherwise quenched in the galaxy, M64
may be in the process of becoming an S0 galaxy. Its surface brightness
profile is in fact remakably similar to that of ESO 383-45, an S0
galaxy also suspected of having suffered a recent merger
\citep{kemp05}. S0 galaxies show antitrunctions more frequently than
other disk types \citep{borlaff14, maltby15}; if mergers drive
evolutionary transition from spirals to S0, they may also lead to
``spheroidal'' antitruncations \citep[denoted Type III-s breaks
  in][]{pohlen06} by growing the galaxy's halo component.

In these cases, however, the halo-like component forming the
antitruncations would by necessity be a different kind of halo than
that surrounding the Milky Way, which appears to have been built up
over time via satellite disruption rather than from heating of the
stellar disk \citep[e.g.][]{morrison00, bullock05, cooper10, ma15}.
Stellar populations in the spheroid beyond the profile breaks would
also appear very similar to those in the disk (where they originated),
which would explain the relatively flat color profiles such
anti-truncated galaxies (including M64) typically exhibit
\citep{zheng15}. Also, if the halo-like component arose due to heating
of the thin disk, and no new thin disk formed from an existing gaseous
disk, the antitrunction would also appear in the mass profile of the
galaxy; this in fact seems to be the case for Type III disks generally
\citep{bakos08, zheng15}.  If merger-spawned spheroids are the root
cause of these antitruncated profiles, such galaxies also should
appear more frequently in dense environments \citep{laine14} either
because of the heightened rate of interactions or simply the
morphology-density relationship raising the likelihood that galaxies
will have significant halo components. However, the fact that M64 is
apparently quite isolated serves to demonstrate that a dense local
environment is not a necessary condition for their formation --- one
merger event may be sufficient.

Finally, the fact that the outer disks of these three galaxies consist
of predominantly old and well mixed populations may simply reflect
their host galaxies' local environment. All three galaxies live in low
density regions --- even the group environment of M106 is sparse, with
no large companion galaxies nearby. Weak interactions with low mass
satellites may not be sufficient to trigger widespread star formation
in outer disks, hence a denser group environment may be more conducive
to triggering outer disk star formation. However, even in denser
groups the evidence is mixed: M101, the dominant galaxy of its
dynamically active group, shows young blue populations in its outer
disk \citep{mihos12, mihos13a}, but in the Leo group, the spirals M95
and M96 both show red outskirts \citep{watkins15}.  This ambiguity is
present in larger surveys as well; \citet{maltby12}, for example,
found little difference in outer disk structure between field and
cluster galaxies, while \citet{erwin12} found significant differences
between field and cluster S0 galaxies (including a complete lack of
disk truncations in cluster S0s).  \citet{roediger12} found that
cluster disk galaxies are distributed equally amongst the three disk
break types, a significant difference compared to field galaxies
\citep{pohlen06}, with significant U-shaped age gradients present in
all three types, in apparent contradiction to the photometric results
of \citet{bakos08}.  Some conflict thus appears to be present regarding
environmental influence on outer disk evolution, which may be
partially resolved if the immediate, local environment is in fact the
driving influence, rather than the global environment.

\section{Summary}

We have performed deep surface photometry (\mub$=$28 - 30) of the nearby
galaxies M94, M64, and M106, and incorporated archival UV and 21cm
\ion{H}{1} data to probe the formation histories of the galaxies' outer
disks. All three galaxies exhibit red outer disks beyond a radius
corresponding to a truncation in star forming activity and high
column-density \ion{H}{1} gas in the disk. A Fourier analysis of the
azimuthal surface brightness and color profiles of each galaxy's outer
disk shows these components are smooth and well-mixed, devoid of spiral
arms or significant non-axisymmetric structure.

New star formation in M94 is truncated at $\sim10$kpc, beyond which
the disk appears azimuthally smooth and red but for some mild
lopsidedness.  The stellar disk, which seems to be continuous despite
the offset inner and outer isophotes, extends to at least $\sim$20kpc,
or $\sim$9 scale lengths, with no emergence of a stellar halo down to
a surface brightness of \mub$\sim$30.  Given M94's isolation and smooth
undisturbed outer disk, our data favors secularly-driven radial
migration of disk populations to explain the galaxy's outer
structure. This, combined with its relatively close distance ($\sim$4
Mpc) makes M94 an ideal test-bed for follow-up studies investigating
how secular evolution processes such as radial migration affect outer
disk formation.

M64 shows a stark star formation truncation only a few kpc from the
center, with a low \ion{H}{1} column density beyond this radius, and a
sharp anti-truncation in the stellar surface brightness beginning
around 400\arcsec \ (9kpc).  We trace this anti-truncated disk to
$\sim$19kpc, or $\sim$13 inner disk scale lengths. M64's strongly
anti-truncated profile is likely the signature of a transition from
the galaxy's disk to its diffuse stellar halo rather than being a true
upbending of the disk surface brightness profile. The recent merger
event in M64 appears to have disrupted its gas disk and truncated star
formation in all but the inner few kiloparsecs, leading to the
galaxy's very flat and red color profile. M64 thus appears to be
undergoing a transition from a spiral to an S0 galaxy, an interesting
example of merger-driven galaxy transformation in an otherwise
isolated environment.

Despite elevated levels of star formation, M106 still shows a clear
star formation truncation radius associated with the end of its outer
spiral arms at $\sim$600\arcsec\ (22kpc).  Its stellar disk extends
roughly twice this distance beyond this truncation radius, with signs
of interaction with its two brightest companion galaxies.  We trace
M106's stellar disk to $\sim$40 kpc, or $\sim$6.5 scale lengths.
Although M106 possesses a more robust satellite system than M64 or
M94, its smooth outer disk and fairly weak tidal structure argues that
these satellites are not dramatically reshaping the disk --- instead,
they may have helped drive the outward migration of stars in M106's
disk without completely disrupting the disk outskirts. M106 may serve
as an interesting comparison to the Mily Way's own satellite-driven
evolution, given the similarity in morphology, luminosity, and local
environment between the two galaxies.

The red colors of these galaxies' outer disks (\bmv$\sim$0.8 in their
outermost regions) imply predominantly old stellar populations. For
exponentially-declining star formation histories, colors this red cannot
be achieved for decay rates longer then $\tau$=2 Gyr; and cannot be
achieved for \emph{any} $\tau$ if metallicities are below [Fe/H]=-0.7.
These properties, along with the smoothness of the outer disks, suggest
that these parts of the galaxies are not formed through on-going or
sporadic star formation, but rather dynamical processes such as heating
or radial migration of stars from the inner disk. The lack of a
significant young stellar population in these galaxies' outskirts may
reflect the sparseness of their local environment; stronger or repeated
encounters may be needed to trigger widespread and sustained star
formation in outer disks. Additional studies of the detailed stellar
populations in outer disks over a wider range of environment would be
informative.

However, while all three of the galaxies studied here do live in low
density environments, they also appear to have different interaction
histories. In this sense, it is interesting that similarly old and
smooth stellar populations exist in the outer disks of each galaxy
irrespective of the influence of their local environments and recent
interaction history --- secular processes that operate in a completely
isolated galaxy (M94) produce a very similar looking outer disk
population as those in a galaxy interacting with companions (M106) or
recovering from a recent merger (M64). Furthermore, the large physical
extent of these azimuthally smooth outer disks implies a very high
efficiency with which stars can be transported via radial migration;
whether such extended disks can be built this way remains unclear.

Finally, while red outer disk colors and U-shaped color profiles are
frequently cited as evidence of radial migration process \citep{bakos08,
martin14, zheng15}, broadband colors leave a great deal of ambiguity
regarding the actual stellar populations producing them. Ambiguity is
present even in many resolved population studies; without a halo field
to compare to, for example, halo star contamination fractions in outer
disk studies remain unconstrained. Measuring stellar kinematics in these
extended regions would be ideal to break the disk/halo ambiguity,
however this remains infeasible for galaxies beyond $\sim$1 Mpc. Until
such studies are possible, combining data from low-resolution, deep
surface photometry (to derive the morphology and integrated
properties of extended regions in galaxies) with resolved star
studies (to deconstruct the detailed stellar populations and star
formation histories of these regions) seems the best option for
future studies of outer disks.

\section{Acknowledgements}

This work has been supported by a Jason J. Nassau Graduate Fellowship
to AEW, and by the National Science Foundation through award 1108964
to JCM.  This work made use of Numpy, SciPy \citep{oliphant07}, and
MatPlotlib \citep{hunter07}.  This work made use of THINGS, `The HI
Nearby Galaxy Survey' \citep{walter08}.  Figures~\ref{fig:m94comp},
\ref{fig:m64comp}, \ref{fig:m106comp}, and \ref{fig:4248} made use
of Min-Su Shin's publicly available code
img\_scale.py\footnote{http://dept.astro.lsa.umich.edu/$\sim$msshin/science/code/Python\_fits\_image/}.
We would also like to thank Ken Duncan for the use of his code SMpy.

{\it Facility:}
\facility{CWRU:Schmidt} - The Burrell Schmidt of the Warner and Swasey
Observatory, Case Western Reserve University

\appendix

\section{ON THE BURRELL-SCHMIDT PSF}

As studies such as this one begin to breach lower and lower surface
brightness limits, concerns about instrumental artifacts become much
more important. Specifically, these concerns focus on the influence of
scattered light, in the form of internal reflections and the extended
wings of the point spread function (PSF), which can skew radial
surface brightness and color profiles at low surface brightness
\citep[for a recent discussion of this problem, see][]{sandin14,
  sandin15}. While aggressive anti-reflection coatings on both the
filters and dewar windows of the Burrell Schmidt minimize bright
reflections in our data, the effect of the extended PSF still remains
at a low level, which we explore in more detail here. Before embarking
on quantitative tests of the influence of the PSF on the derived
photometric profiles, we first note that we have seen no evidence of
systematic reddening of galaxy profiles in previous studies using the
Burrell Schmidt. While the three galaxies studied here show red outer
disks, Schmidt imaging of the spiral galaxy M101 \citep{mihos13a}
revealed blue outer isophotes, while deep imaging of the Virgo Cluster
showed blueward gradients in the diffuse outer halos of the massive
ellipticals M87 \citep{rudick10} and M49 \citep{mihos13b}. Redward
gradients therefore do {\it not} seem to be a systematic result of our
imaging techniques. However, to assess this effect more
quantitatively, in this Appendix we convolve an updated and more
accurate measurement of the Burrell Schmidt PSF with a variety of
galaxy profiles to quantify its effect on the extracted surface
brightness and color profiles of our galaxies.

\begin{figure*}
  \centering
  \includegraphics[scale=0.57]{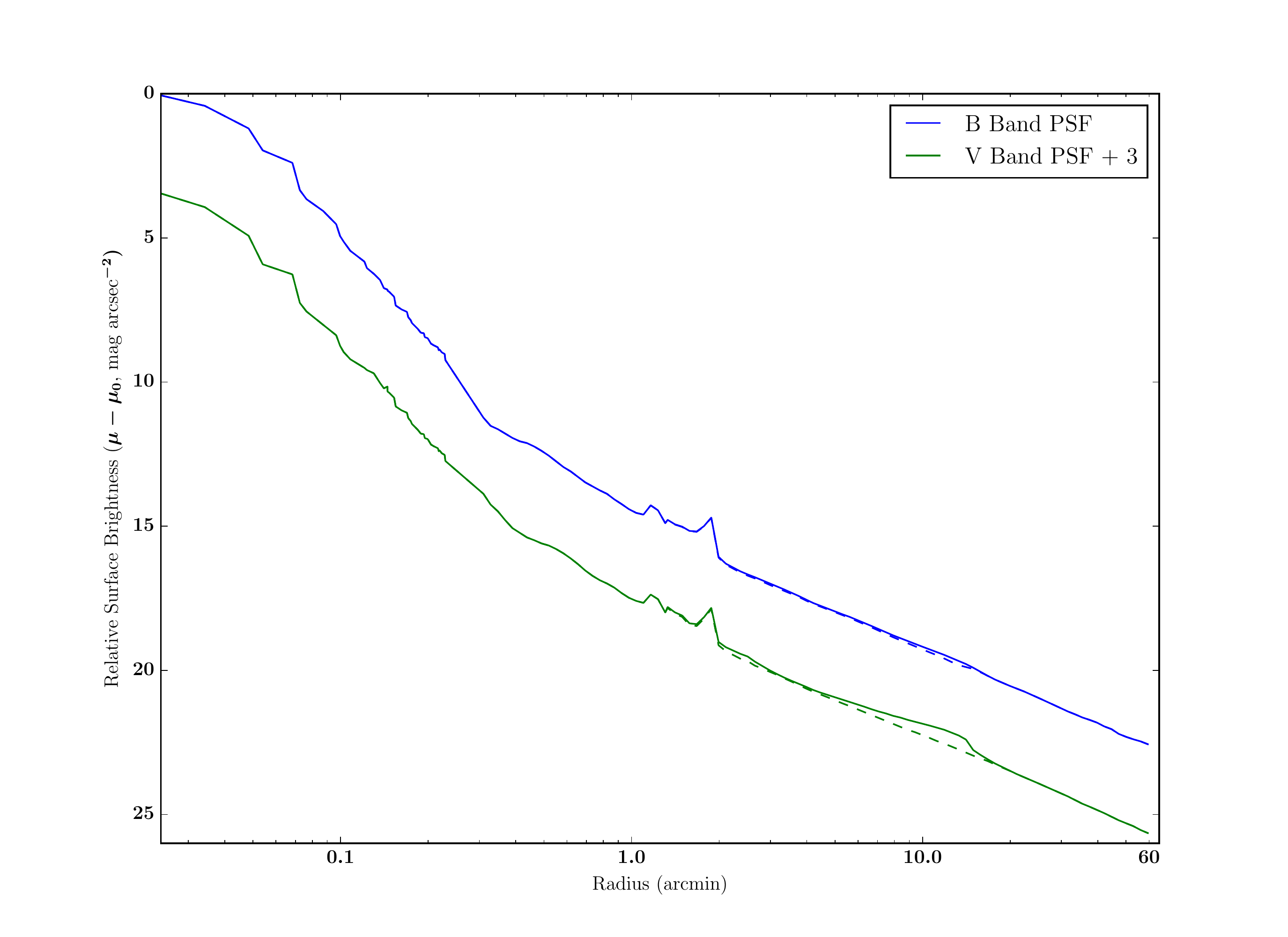}
  \caption{The \emph{B} Band and \emph{V} Band (offset by 3 magnitudes
    for clarity) radial profiles of the Burrell Schmidt PSF.  Solid
    lines show the profiles including reflections, as measured from
    our bright star exposures, while dashed lines show the underlying
    profile wings (with reflections subtracted out).  Profiles are
    normalized such that $\mu_{0} = 0$.
    \label{fig:psf}}
\end{figure*} 

As discussed briefly in Section~2.1, we measured the Burrell Schmidt PSF
using long exposures of bright stars in order to subtract bright
reflection halos and the extended PSF wings around bright stars. We
show the Burrell Schmidt PSF radial profile in the \emph{B} and \emph{V}
bands in Figure \ref{fig:psf} out to one degree. An earlier measurement
of the \emph{V} band PSF radial profile was published in
\citep[hereafter S09]{slater09} and \citep{janowiecki10}; in the outer
wings, this more recent measurement compares well to the older profile,
despite being taken several years later and using a different CCD. We
note, however, that the inner core profile shown in Figure \ref{fig:psf}
does differ significantly from that of S09. That earlier study focused
on proper subtraction of the outer wings of the PSF, which is
insensitive to the shape of the inner core. As such, the core profile of
S09 was largely illustrative and not well-determined. In the present
study we have worked to produce a much more accurate measurement of the
PSF core ($r \lesssim$10\arcsec) by using four stars of different
brightnesses to ensure that all pieces of the profile link up correctly.
Our updated profile shows that the core profile illustrated in S09 had
actually been underestimated significantly. While this difference has
little effect on scaled {\it subtraction} of the outer PSF, it has a dramatic
effect on the normalized PSF used for image {\it convolution}. This is simply
due to the fact that there is a range of $\sim$22 magnitudes in
brightness between the core and the PSF wings at $r \sim$1\arcdeg, hence
most of the flux is contained within the core. As such, an error in
the core profile can create significant variation in the intensity of
the wings after normalization.

With a more accurate PSF measurement in hand, we tested its
influence on the derived photometric profiles by convolving several
model galaxies with our normalized PSFs in both bands, and measuring the
resulting surface brightness and color profiles. The model galaxies were
constructed to represent idealized versions (smooth exponential disks of
constant \bmv\ color, with bulges of a constant redder color) of M106,
M64, and M101, in order to test the influence of the galaxies' angular
sizes and central surface brightnesses. The M106 and M64 models had
similar values of $\mu_{0}$, but different scale lengths (see
Figures~\ref{fig:m64profs} and \ref{fig:m106profs}), while the M101
model had a much lower value of $\mu_{0}$ and large angular size
\citep[M101 is nearly face-on; see][]{mihos13a}.

We found that for the M106 and M64 facsimiles, the PSF induces a color
change of $\Delta$\bmv$\sim +0.1$ by a surface brightness of
\mub$\sim$28.0. This change in color occurs beyond where our photometry
is noise-limited by 0.5 mags arcsec$^{-2}$; brighter than this surface
brightness, the color change induced by the PSF is much smaller than
that seen in the data. For example, between \mub$\sim$25.5 and 26.5,
M106 shows a color change of 0.08 magnitudes, while the convolved model
galaxy shows a change of only 0.015 magnitudes. Thus, while some of the
redward gradient in these galaxies' outer disks may be attributable to
the PSF, it is clear that most of the gradient is attributable to
changing stellar populations. It should also be noted that, despite its
relatively smaller angular size, we see no evidence that the PSF is
inducing the anti-truncation seen in M64's outer disk; a significant
PSF-induced anti-truncation is only seen in the convolved model of M64
beyond \mub$\sim$30. Finally, in the M101 facsimile, we see the same 0.1
magnitude color change setting in, but at a much lower surface
brightness of \mub$\sim$30.0. This is simply due to M101's lower {\it
central} surface brightness, which scatters less light to large radius
in the PSF. Taken as a whole, the results of these various tests thus
show that the scientific results presented in this paper (and in
previous papers using data taken with the Burrell Schmidt) are robust to
PSF influence; the error budget is dominated by photometric
uncertainties quantified in Section~2 and 3.

\end{document}